\documentclass[a4paper,11pt,aps,prx,amsmath,amssymb,amsfonts,superscriptaddress,notitlepage,reprint,longbibliography, nofootinbib]{revtex4-2}

\usepackage{graphicx} 
\usepackage{lipsum}
\usepackage{color} 
\usepackage{lmodern} 
\usepackage{textcomp} 
\usepackage[T1]{fontenc} 
\usepackage{siunitx}
\usepackage{mathtools}
\usepackage{xcolor}

\newcommand*{\ket}[1]{|{#1}\rangle}
\newcommand*{\bra}[1]{\langle{#1}|}

\newcommand{\creat}[1]{#1^\dagger}
\newcommand{\annih}[1]{#1}
\newcommand{\derivative}[2]{\frac{\mathrm{d}  #1}{\mathrm{d}#2}}
\def\be{\begin{equation}}
\def\ee{\end{equation}}
\def\bes{\begin{equation*}}
\def\ees{\end{equation*}}

\newcommand{\wsq}{\omega_\mathrm{SQ}}
\newcommand{\chisq}{\chi_\mathrm{SQ}}

\definecolor{myrefcolor}{RGB}{200,20,20} 
\definecolor{mycitecolor}{RGB}{34, 139, 34}   

\usepackage[
    colorlinks=true,      
    linkcolor=myrefcolor, 
    citecolor=mycitecolor,
    urlcolor=magenta,     
    filecolor=violet      
]{hyperref}

\newcommand{\subautoref}[2]{\hyperref[#1]{\autoref*{#1}~\textbf{#2}}}

\usepackage{soul}
\sethlcolor{orange!40}

\begin{document}

\title{Quantum Acoustics with Tunable Nonlinearity in the Superstrong Coupling Regime}

\author{Marco Scigliuzzo}
\thanks{These two authors contributed equally}
\affiliation{Center for Quantum Science and Engineering, \\ \'{E}cole Polytechnique F\'{e}d\'{e}rale de Lausanne (EPFL), CH-1015 Lausanne, Switzerland}
\affiliation{%
Laboratory of Photonics and Quantum Measurements (LPQM),  Institute of Physics, EPFL, CH-1015 Lausanne, Switzerland
}%
\affiliation{Department of Microtechnology and Nanoscience, Chalmers University of Technology, 412 96 Gothenburg, Sweden}
\author{Léo Peyruchat} 
\thanks{These two authors contributed equally}
\affiliation{Hybrid Quantum Circuits Laboratory (HQC), Institute of Physics, \'{E}cole Polytechnique F\'{e}d\'{e}rale de Lausanne (EPFL), 1015 Lausanne, Switzerland}
\affiliation{Center for Quantum Science and Engineering, \\ \'{E}cole Polytechnique F\'{e}d\'{e}rale de Lausanne (EPFL), CH-1015 Lausanne, Switzerland}
\author{Riccardo Maria Marabini}
\affiliation{Hybrid Quantum Circuits Laboratory (HQC), Institute of Physics, \'{E}cole Polytechnique F\'{e}d\'{e}rale de Lausanne (EPFL), 1015 Lausanne, Switzerland}
\affiliation{Center for Quantum Science and Engineering, \\ \'{E}cole Polytechnique F\'{e}d\'{e}rale de Lausanne (EPFL), CH-1015 Lausanne, Switzerland}
\author{Carla Becker}
\affiliation{Hybrid Quantum Circuits Laboratory (HQC), Institute of Physics, \'{E}cole Polytechnique F\'{e}d\'{e}rale de Lausanne (EPFL), 1015 Lausanne, Switzerland}
\affiliation{Center for Quantum Science and Engineering, \\ \'{E}cole Polytechnique F\'{e}d\'{e}rale de Lausanne (EPFL), CH-1015 Lausanne, Switzerland}
\author{Vincent Jouanny}
\affiliation{Hybrid Quantum Circuits Laboratory (HQC), Institute of Physics, \'{E}cole Polytechnique F\'{e}d\'{e}rale de Lausanne (EPFL), 1015 Lausanne, Switzerland}
\affiliation{Center for Quantum Science and Engineering, \\ \'{E}cole Polytechnique F\'{e}d\'{e}rale de Lausanne (EPFL), CH-1015 Lausanne, Switzerland}
\author{Per Delsing}
\email[Corresponding author: ]{per.delsing@chalmers.se}
\affiliation{Department of Microtechnology and Nanoscience, Chalmers University of Technology, 412 96 Gothenburg, Sweden}
\author{Pasquale Scarlino}
\email[Corresponding author: ]{pasquale.scarlino@epfl.ch}
\affiliation{Hybrid Quantum Circuits Laboratory (HQC), Institute of Physics, \'{E}cole Polytechnique F\'{e}d\'{e}rale de Lausanne (EPFL), 1015 Lausanne, Switzerland}
\affiliation{Center for Quantum Science and Engineering, \\ \'{E}cole Polytechnique F\'{e}d\'{e}rale de Lausanne (EPFL), CH-1015 Lausanne, Switzerland}

\date{\today}

\begin{abstract}
Precise control of mechanical modes in the quantum regime is a key resource for quantum technologies, offering promising pathways for quantum sensing with macroscopic systems and scalable architectures for quantum simulation.
In this work, we realise a multimode mechanical cavity coupled to a superconducting Kerr resonator, which induces nonlinearity in the mechanical modes.
The Kerr mode is realised by a flux-tunable SQUID array resonator, while the mechanical modes are implemented by a surface acoustic wave (SAW) cavity.
Both mechanical and electromagnetic modes are individually addressable via dedicated measurement lines, enabling full spectroscopic characterisation.
We introduce a straightforward protocol to measure the SQUID array resonator's participation ratio in the hybrid acoustic modes, quantifying the degree of hybridisation.
The participation ratio reveals that our device operates at the onset of the multimode coupling regime, where multiple acoustic modes simultaneously interact with the nonlinear superconducting element.
Furthermore, this platform allows controllable Kerr-type nonlinearities in multiple acoustic modes, with the participation ratio serving as the key parameter determining both the dissipation rates and nonlinear strengths of these hybridised modes.
Close to the resonant regime, we measure a cross-Kerr interaction between seven pairs of mechanical modes, which is controllable via the SQUID array resonator detuning.
These results establish a platform for engineering nonlinear multimode mechanical interactions, offering potential for future integration with superconducting qubits and implementation of multiple mechanical qubits.
\end{abstract}

\maketitle

\section{Introduction}

Developing the quantum control of mechanical modes represents a fundamental task for quantum science and technologies. 
Due to their mass, mechanical systems possess distinctive properties, such as sensitivity to weak external forces, acceleration, and even gravitational effects \cite{barzanjeh2022optomechanics, abbott2016observation}.
Moreover, the slow propagation speed of mechanical excitations results in a reduced footprint for the mechanical system and enables exploration of exotic interaction regimes, such as the "giant atom" limit \cite{gustafsson2014propagating, andersson2019non}.

Over the past two decades, significant advances have established the field of quantum acoustics \cite{POOT2012273, clerk2020hybrid}, with a wide range of hybrid platforms developed to interface mechanical degrees of freedom with quantum systems. These include levitated particles \cite{ashkin1986observation}, cold atoms \cite{phillips1982laser}, and a variety of solid-state systems \cite{o2010quantum}. Among these, solid-state mechanical platforms stand out by leveraging superconducting circuit technologies and lithographically defined mechanical elements, such as phononic crystal defects \cite{arrangoiz2018coupling}, membranes \cite{teufel2011sideband, kotler2021direct}, bulk acoustic resonators~\cite{chu2017quantum}, and surface acoustic wave (SAW) devices \cite{gustafsson2014propagating, manenti2017circuit, satzinger2018quantum, sletten2019resolving, noguchi2017qubit, chen2025chip, hugot2025approaching}.

These systems have enabled remarkable control over single mechanical modes, including the preparation of squeezed states \cite{wollman2015quantum, marti2024quantum}, Fock states \cite{chu2018creation, qiao2023splitting}, and other nonclassical states \cite{bild2023schrodinger}. In particular, introducing nonlinearity into mechanical modes has proven crucial for enabling their direct quantum control \cite{yang2024mechanical}.
However, despite these achievements, progress has been largely confined to single-mode mechanical systems. While a few efforts have aimed to extend quantum control to multiple mechanical modes \cite{li2025quantum}, true multimode quantum control remains an outstanding challenge.

In this context, SAW resonators offer a distinct advantage: their acoustic mode confinement is restricted to a well-defined frequency band. As a result, electromagnetic resonators or qubits can be tuned outside this range without interacting with the SAW modes, enabling greater flexibility in device design. This frequency stop band is lithographically defined by the Bragg gratings during fabrication~\cite{manenti2016surface}.
The multimode nature of SAW resonators has been leveraged to probe high-density ensembles of surface two-level systems (TLSs) \cite{andersson2021acoustic, gruenke2025surface}, providing a valuable tool for material and interface characterisation. 
In addition, SAW devices are highly versatile and readily integrable with a broad range of platforms, including optical cavities~\cite{tadesse2014sub}, photonic waveguides~\cite{beugnot2014brillouin}, and even propagating optical photons~\cite{iyer2024coherent}. Their compatibility extends further to solid-state quantum systems, having been successfully coupled to vacancy centers in various substrates~\cite{whiteley2019spin, maity2020coherent} and to electrically defined semiconductor qubits~\cite{sato2017two, schutz2017universal}.

In this work, we investigate the onset of the multimode (i.e. superstrong) coupling regime~\cite{meiser2006superstrong,sundaresan2015beyond,moores2018cavity,puertas2019tunable,kuzmin2019superstrong} between mechanical modes of a SAW cavity and a flux-tunable, nonlinear electromagnetic ancilla~\cite{gao2018programmable,ma2021quantum}. The system consists of 29 mechanical modes with distinct frequencies $\omega_i$, each coupled to the fundamental mode of a superconducting SQUID array resonator, whose frequency $\wsq$ can be tuned in situ via external magnetic flux.

We begin by performing single-tone spectroscopy to establish the multimode coupling regime, characterised by the condition where the mechanical mode frequency spacing (free spectral range, $\Delta\omega_i$) becomes comparable to the individual coupling strengths $g_i$ between the electromagnetic and mechanical modes.
To quantify the hybridisation between the electromagnetic and mechanical degrees of freedom, we introduce a participation-ratio metric based on the relative frequency shifts of the mechanical modes as a function of the SQUID array resonator detuning. This metric enables accurate prediction of both the dissipation and nonlinear response of the hybridised modes.
We further experimentally observe and characterise cross-Kerr interactions $\chi_{ij}$ among seven distinct acoustic modes, demonstrating the emergence of nonlinear interactions mediated by the shared coupling to the nonlinear electromagnetic ancilla. These results confirm the nonlinear multimode character of the system and validate the participation-based model.
Finally, we discuss how this platform can be extended toward implementing multiple mechanical qubits~\cite{yang2024mechanical}.

\section{Results}

\subsection{Setup and Device}

\begin{figure}
    \includegraphics[width = \linewidth]{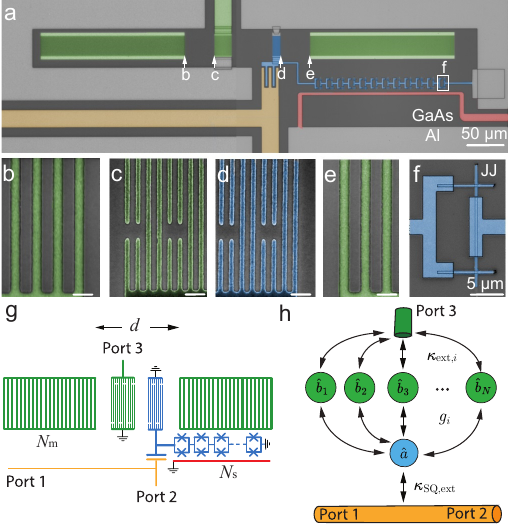}
    \caption{\label{fig:concept}
    \textbf{Surface acoustic wave cavity coupled to a SQUID array resonator} 
    (a) False color optical image of the device.   
    (b,e) Two Bragg mirrors form an acoustic cavity.
    (c) A large launcher IDT with 30 pairs of split fingers is coupled to a microwave feedline to probe SAW modes in reflection.
    (d) A SQUID array forms a nonlinear $\lambda/4$ resonator with a shunt to ground on the right side, and a capacitive end on the left side which connected to a microwave feedline in notch configuration (yellow) and to SAW modes via a smaller coupler IDT with 14 pairs of split fingers.
    A flux line (red) enables uniform flux threading.
    (f) SEM micrograph showing an enlargement of the highlighted white rectangle in panel a, reporting a zoom-in of a single SQUID. 
    (g) Circuit schematic of the systems. 
    The mirrors are separated by $d=\SI{155}{\um}$ and both present a Bragg grating with $N_m=500$ finger elements, much longer than the penetration length. 
    The SQUID array resonator features $N_s=12$ SQUID units.
    (h) Mode diagram representing the SAW cavity modes, $\hat b_i$, coupled to the SQUID array resonator mode,~$\hat a$, together with their respective external dissipation rates, $\kappa_{\mathrm{ext},i}$ and $\kappa_{\mathrm{SQ,ext}}$.
    }
\end{figure}

\textbf{Device:}
The device presented in \subautoref{fig:concept}{} features a nonlinear electromagnetic resonator coupled to a linear multimode mechanical cavity. 
The electromagnetic resonator is implemented as an interdigital capacitor shunted by an array of $N_\mathrm{s}=12$ SQUIDs.
The frequency $\wsq(\phi_X)$ of the fundamental mode $\hat a$ can be tuned by threading magnetic flux $\phi_X$ through each SQUID loops. 
The additional $N_{\rm s}-1$ modes have a much higher frequency, and they do not impact the physics investigated in this work.
The mechanical modes $\hat b_i$ are SAWs confined between two Bragg mirrors and directly excited or detected by an interdigital transducer (launcher IDT). 
The coupling between the electromagnetic and mechanical excitations is implemented by an interdigital transducer at the capacitive end of the SQUID array resonator (coupler IDT). 

The SAW components are designed for wavelength $\lambda_0=\SI{720}{\nm}$.
A Bragg mirror consists of $N_{\rm m}=500$ fingers \SI{180}{\nm} wide realised with \SI{30}{\nm} thin aluminium films (see \subautoref{fig:concept}{b, e}) that yield  $\approx \SI{100}{\MHz}$ wide stopband around the frequency $v_{\rm SAW}/\lambda_0\approx$ \SI{4}{\GHz}. 
The distance between the mirrors is $d=\SI{155}{\micro\meter}$.
Both the launcher and coupler IDTs (see \subautoref{fig:concept}{c, d} respectively)  are realised with split finger geometry to prevent internal reflection within the mirror \cite{morgan2010surface}, and they have  $30 \ \text{and} \ 14$ pairs of split fingers, respectively.

The sample is bonded on a PCB, enclosed in a trilayer Cu-Al-$\mu$-metal shield, mechanically and thermally anchored at the mixing chamber plate of a dilution refrigerator with base temperature of \SI{10}{\milli\kelvin} (see \autoref{sec:appendix_meas_setup} for more information). 

As depicted in the circuit schematic of \subautoref{fig:concept}{g}, the resonator is coupled to the microwave feedline in a notch configuration and can be measured via ports 1 and 2; similarly, the SAW modes can be directly excited and measured in reflection from port 3. 
The mode diagram is finally reported in \subautoref{fig:concept}{h}.
Throughout the article, we use blue colors for $S_{21}$ and green for $S_{33}$ scattering coefficients.

\subsection{Bare Elements}

The Hamiltonian of the device in the bare mode basis is 
\begin{equation}
\begin{aligned}
    H/ \hbar =  \; & \omega_\mathrm{SQ}(\phi_X) \, \creat{a}a + \chi_{\mathrm{SQ}} (\creat{a}a)^2+ \sum_{i=1}^{29} \omega_i \, b_i^{\dagger} b_i 
    \\ &+ \sum_{i=1}^{29} g_i (a^{\dagger} b_i +  a b_i^{\dagger}) ,
\label{eq:H_bare_rwa}
\end{aligned}
\end{equation}
where $\chisq$ is the SQUID array resonator nonlinearity, $\omega_i$ is the resonance frequency of the $i$-th SAW cavity mode, and $g_i$ the coupling between $i$-th SAW cavity and SQUID array resonator (see \subautoref{fig:concept}{h}).

\begin{figure}
    \includegraphics[width = \linewidth ]{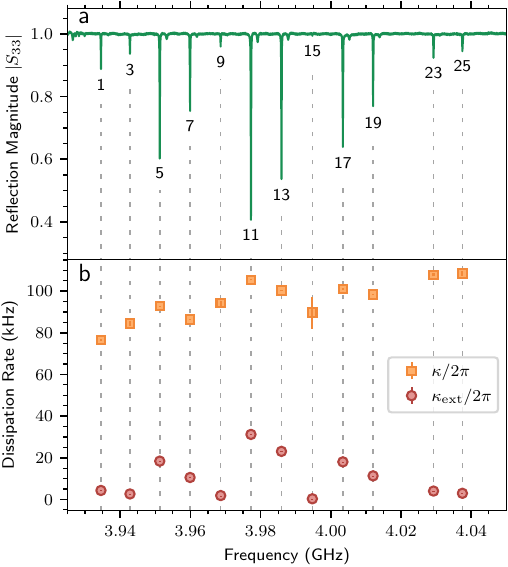}
    \caption{\label{fig:bare_saw}\textbf{Bare SAW modes.}
    (a) Reflection magnitude $|S_{33}|$ measured from the SAW feedline by the launcher IDT.
    Dashed lines and index highlight the frequency of longitudinal SAW modes.
    (b) Extracted dissipation rates of the SAW modes. Longitudinal SAW modes are undercoupled: the total dissipation rate (orange square) is larger than twice the external one (red circle).
    Small dips are transverse SAW modes with smaller external coupling rates and larger internal dissipation rates.
    We used a probe power of $-131$\,dBm at device level, leading to an average phonon population of $23$ across the longitudinal SAW modes.
    Error bars are obtained from the fit and represent one standard deviation.    
    }
\end{figure}

\textbf{Bare SAW modes:}
We first characterise the bare SAW modes by setting the SQUID array resonator at $\wsq(0)/2\pi \approx \SI{4.7}{\GHz}$, corresponding to \SI{700}{\MHz} detuning from the central frequency of the SAW cavity, and measuring $S_{33}$ reported in \subautoref{fig:bare_saw}{a}.
We extract the slowly varying background of $S_{33}$ to normalise the data.
The index number reported in the figure represents the longitudinal SAW modes \cite{morgan2010surface}, whose frequencies range from \SI{3.93}{\GHz} to \SI{4.05}{\GHz}, with a free spectral range of approximately \SI{8.6}{\MHz}.
The dissipation rates of each SAW mode are fitted to \autoref{Eq:reflection_coeff_single_mode} and reported in \subautoref{fig:bare_saw}{b}.
The logitudinal SAW modes are undercoupled to the launcher IDT, with approximately constant $\kappa_\mathrm{int} /2\pi \approx $ \SI{80}{\kHz} internal dissipation rate for all modes.
The maximum external coupling $\kappa_\mathrm{ext}/2\pi \approx$ \SI{40}{\kHz} is achieved for mode 11 near the center of the mirror bandwidth. 
The measurement is performed with $-132$ dBm input power at device level, leading to an average occupation of 23 phonons (see \autoref{Eq:Photon_number_conversion} for calibration). 
For a more general characterisation of the dissipation rates as a function of probe power, see \autoref{fig:saw_dissip_power_appendix} in appendix.
Due to the spatial profile of the mechanical modes, some modes are weakly coupled to the launcher IDT (e.g. mode 21 expected at \SI{4.02}{\GHz}) \cite{sletten2019resolving}.

We also observe transverse SAW modes \cite{fisicaro2025imaging}, weakly coupled to the launcher IDT ($\kappa_\mathrm{ext}/2\pi < \SI{5}{\kHz}$) and with larger internal losses ($\kappa_\mathrm{int}/2\pi \approx \SI{315}{\kHz}$).
The position and dissipation of transverse modes are detailed in \autoref{fig:saw_dissip_appendix}.

\begin{figure*}
    \includegraphics[width = \textwidth ]{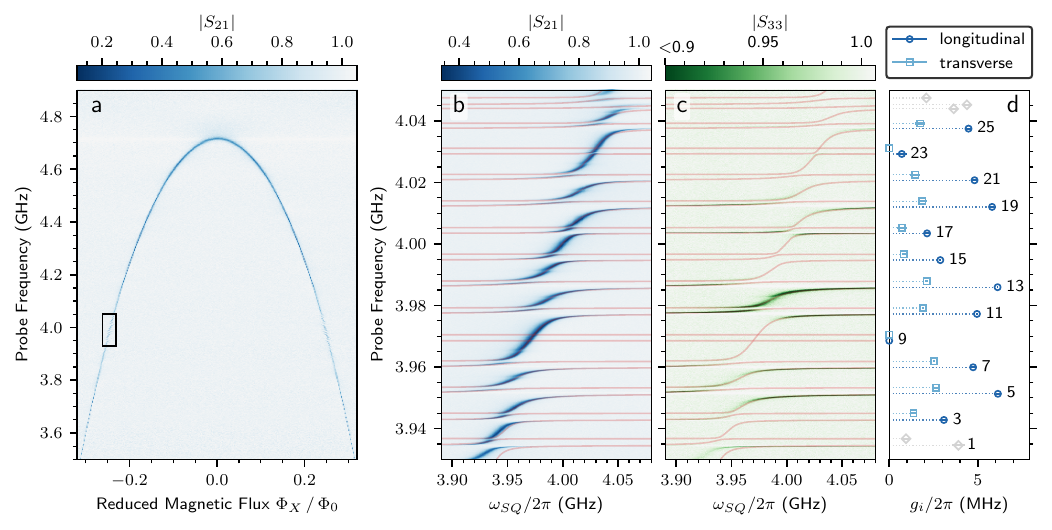}
    \caption{\label{fig:fluxsweep}\textbf{Flux sweep of the SQUID array resonator across SAW modes.}
    (a) Normalised transmission $S_{21}$ through the SQUID array resonator feedline, measured as a function of flux threading the SQUID array generated by an external coil.
    The power of the probe tone is \SI{-142}{\dB}m.
    (b,c) Magnitude of $S_{21}$ (blue) and $S_{33}$ (green) measured as a function of flux sweeps across the SAW modes in the region highlighted by the black rectangle in panel (a).
    The power of the probe tone applied to port 1 is \SI{-132}{\dB}m, while the power of the probe tone applied to port 3 is \SI{-131}{\dB}m.
    The peak position obtained from a multimode fit
    is shown with red solid lines, and is in excellent agreement with the experimental data except for the modes at the edge of the mirror bandwidth (grey in (d)).
    (d) Coupling rates between the electromagnetic and mechanical modes, $g_i$, extracted from the multimode fit.
    The couplings to longitudinal SAW modes are shown with dark blue circle markers, while the couplings to transverse SAW modes are shown with light blue square markers.
    Error bars are obtained from the fit and represent one standard deviation.    
    }
\end{figure*}

\textbf{Bare SQUID array resonator:}
We now characterise the bare SQUID array resonator.
In \subautoref{fig:fluxsweep}{a}, we measure the transmission $S_{21}$ of the waveguide (see \autoref{fig:concept}) while sweeping the flux in the SQUIDs by threading current in an external superconducting coil.
The transmission data is normalised by a background trace measured with the SQUID array tuned at low frequency (near half flux quanta).
We extract the SQUID array resonator frequency and fit it to \autoref{eq:SQUID_characterisation} to convert applied current in the external coil to bare SQUID array resonator frequency $\wsq$.

The external coupling rate $\kappa_\mathrm{SQ,ext}$ to the feedline scales with $\wsq^2$, and remains in the range 4-\SI{6}{\MHz} (see the fit to \autoref{fig:SQ_dissip_appendix} in the appendix).
The internal dissipation rate at single photon is limited by the electric field participation ratio in the launcher IDT, except in the frequency region approximately \SI{300}{\mega\hertz}-wide where the IDT coupler emits SAWs (see \autoref{app:SQUID_array_dissipation} for more details).

The direct coupling from the SQUID feedline to SAW modes via the coupler IDT is negligible, with external coupling rates below \SI{0.4}{\kilo\Hz}.

\subsection{SAW-SQUID hybridisation}

\textbf{Flux sweep and resonant coupling:}
In \subautoref{fig:fluxsweep}{b, c}, we report a detailed sweep of the flux in the range where the bare SQUID array resonator frequency is within the SAW mirror stopband, and measure the response of the SQUID array resonator mode $\hat a$ ($S_{21}$), and the SAW modes $\hat{b}_i$ ($S_{33}$), respectively. 
The extinction rates of these resonances differ in the two measurements due to different coupling to the relative excitation ports (notice that the colour bar legend shows different amplitudes for each panel). 

Combining the information in panels b and c, we count 29 SAW modes. 
However, in the characterisation shown in \autoref{fig:bare_saw}, we only found the frequency of 12 longitudinal modes and 6 transverse SAW modes. 
Therefore, we fit the detailed flux sweeps to the linear part of Hamiltonian in \autoref{eq:H_bare_rwa} with the frequency of the 11 missing SAW modes and all the 29 $g_i$ couplings as free parameters.
The couplings $g_i$ are different for each SAW mode due to the different overlap of their mode shape with coupler IDT; the values obtained from the fit are shown in \subautoref{fig:fluxsweep}{d}. 
The coupling to transverse SAW modes is about three times weaker than to longitudinal modes due to lower overlap with the coupler IDT.

The frequency difference $\Delta \omega_i = \omega_{i+1} - \omega_i$ between SAW modes alternates between approximately 2.0 and $\SI{6.5}{\MHz}  \times  2\pi$, as shown in \autoref{fig:saw_fsr_appendix}.
The couplings $g_i$ are comparable to $\Delta \omega_i$, with on average $g_\mathrm{longitudinal} /2\pi \approx3.8$ MHz and $g_\mathrm{transverse} /2\pi \approx1.4$ MHz.
The device is thus at the onset of the multimode (or superstrong) coupling regime with $g_i \gtrsim \Delta \omega_i$~\cite{meiser2006superstrong, sundaresan2015beyond, moores2018cavity, puertas2019tunable, kuzmin2019superstrong}, the consequence of which we will study in the rest of the article.
More details on the fitting procedure are given in \autoref{subsec:appendix_multimode_fit}.

\textbf{Model in hybridised mode basis:}
To study the system in the resonant regime of mechanical and electromagnetic modes ($\omega_\mathrm{SQ} \approx \omega_i$), it is convenient to rewrite the Hamiltonian from \autoref{eq:H_bare_rwa} in the basis of hybridised modes.
As detailed in \autoref{sec:bogoliubov_transformation}, we apply a Bogoliubov transformation \cite{del2004quantum} to the linear part of the Hamiltonian and expand the Kerr term perturbatively to obtain
\begin{align}
    \tilde{H}/\hbar = \sum_{i=0}^{29} \tilde{\omega}_i \, \creat{c_i}\annih{c_i} + \sum_{i,j=0}^{29} \tilde{\chi}_{ij} \, \creat{c_i}\creat{c_j}\annih{c_i}\annih{c_j}\;.
    \label{eq:H_hybrid_rwa}
\end{align}
All the hybridised modes $c_i$ now have self-Kerr $\tilde \chi_{i,i}$ and cross-Kerr $\tilde \chi_{i \neq j}$ nonlinearities.
The hybridised modes $c_i$ can be decomposed in the bare mode basis as 
\begin{equation}
    c_j = a \,u_{\mathrm{SQ},j} + \sum_{i=1}^{29} b_i \,u_{i,j} \  ,
\end{equation}
where $u_{i,j}$ corresponds to the coefficients defined in \autoref{Eq:transformation_a_to_c_bosons}, and $u_{\mathrm{SQ}, j} = u_{0,j}$.
As we demonstrate below, the nonlinearities $\tilde \chi_{i,j}$ and dissipation rates $\tilde \kappa_i$ of the $c_i$ modes 
can be simply obtained from the participation of the SQUID array resonator $|u_{\mathrm{SQ},i}|^2$.

\begin{figure*}
    \includegraphics[width = \linewidth ]{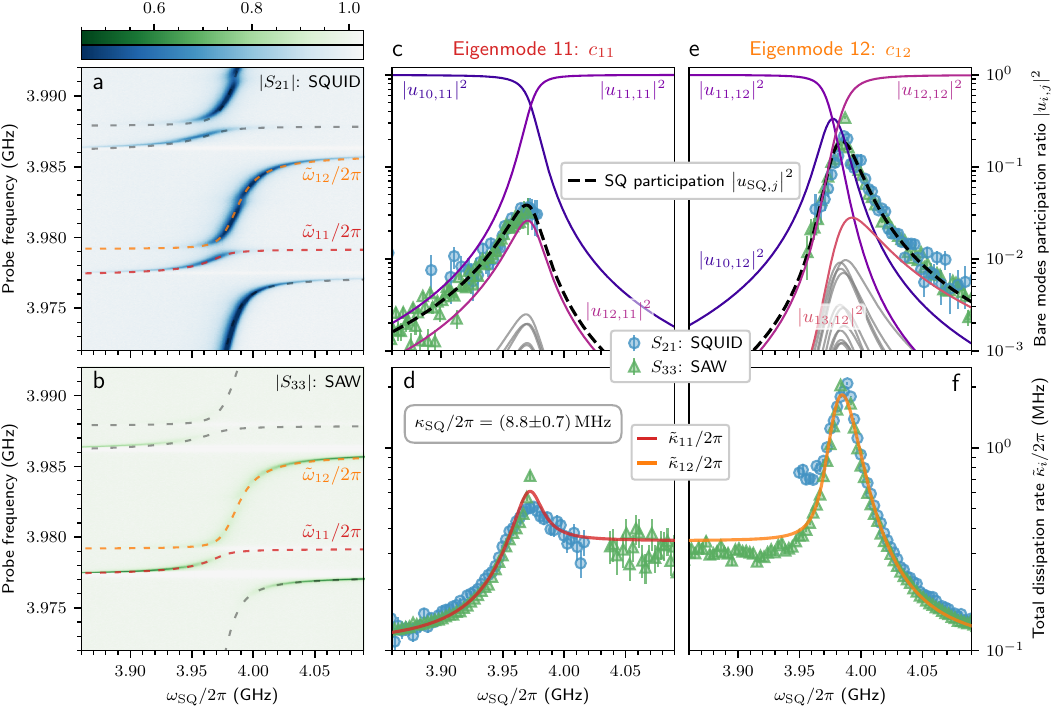} \caption{\label{fig:participation}\textbf{Characterisation of the multimode coupling regime.}
    (a,b) Enlarged view of the magnitude of $S_{21}$ (blue) and $S_{33}$ (green), as shown in \subautoref{fig:fluxsweep}{b,c}, highlighting four selected SAW modes. 
    Dashed lines show the extracted hybridised modes position, with eigenmodes 11 and 12 highlighted in red and orange respectively.
    (c,e) Participation of the bare modes in the eigenmodes $c_{11}$ (c) and $c_{12}$ (e).
    The values of $|u_{\mathrm{SQ},i}|^2$ obtained from $S_{21}$ are shown with blue circle markers, while those obtained from $S_{33}$ are shown with green triangle markers.
    The continuous coloured lines represent the participation ratios of the bare SAW modes, while the black dashed line indicated the participation of the SQUID array resonator mode, computed from numerical diagonalisation of the linear Hamiltonian in \autoref{eq:H_bare_rwa}.  
    Modes with maximum participation larger than $2 \%$ are shown in color, while other SAW modes with lower participation are shown in grey.
    (d,f) Total dissipation rates of eigenmodes $c_{11}$ (d) and $c_{12}$ (f), extracted by fitting each individual resonance to the  input-output relations \autoref{Eq:Notch_coefficient_single_mode_FANO} ($S_{21}$) or \autoref{Eq:reflection_coeff_single_mode} ($S_{33}$).
    The resulting total dissipation rates from $S_{21}$ and $S_{33}$ are shown as blue circles and green triangles, respectively.
    Solid lines show a global fit to \autoref{eq:hybrid_kappa}, from which we extract total dissipation rate of the SQUID array resonator inside the mirror stopband, $\kappa_\mathrm{SQ}/2\pi = 8.8 \pm 0.7$ MHz. 
    All marker error bars represent one standard deviation.
}
\end{figure*}

\textbf{Participation ratio analysis:}
The participation ratio of nonlinear circuit elements is a widely used concept for modelling and designing superconducting circuits \cite{nigg2012black, minev2021energy}. 
This method provides an efficient way to analyze systems combining linear and nonlinear elements. 
Unlike a full quantum simulation of the device, this approach simplifies the problem by first performing a linear simulation to calculate the eigenmodes of the system. 
The system's couplings and nonlinearities are then determined from the overlap between the eigenmodes and the nonlinear elements.

Here, we propose a simple and accurate method to extract the SQUID array resonator participation directly from measurements of the hybrid mode frequency as a function of the SQUID array resonator frequency. Specifically, we derive in the Appendix (\autoref{Eq:Participation_from_derivative}) the following relationship:
\begin{equation}
    |u_{\mathrm{SQ},i}|^2 = \derivative{ \tilde \omega_i }{ \omega_\mathrm{SQ}}.
    \label{eq:particip_ratio_deriv}
\end{equation}

\textbf{Participation ratio measurement from eigenmode derivative:}
Two coupled resonant systems hybridise forming superpositions of the two bare states.
However, because our system is in the multimode coupling regime, the bare SQUID array resonator mode $\hat a$ hybridises simultaneously with multiple bare SAW modes $\hat b_i$. 
As a result, the participation of the SQUID array resonator in the hybrid eigenmodes remains small through the avoided crossings. 
While this feature has been discussed in previous works \cite{moores2018cavity}, it has not been explicitly measured.
We now apply our proposed method (\autoref{eq:particip_ratio_deriv}) to directly measure the SQUID array resonator participation in hybrid modes $ c_{11}$ and $c_{12}$ and verify these predictions.

In \autoref{fig:participation}, we fit the measurements of the detailed flux sweeps in panels a and b to the input output model of \autoref{Eq:Notch_coefficient_single_mode_FANO} for $S_{21}$ and \autoref{Eq:reflection_coeff_single_mode} for $S_{33}$, and obtain the frequencies and dissipation rates of the hybridised modes.
We numerically compute the derivative of the measured frequencies as a function of $\omega_\mathrm{SQ}$.
Blue and green markers in panels (c) and (e) show the derivative obtained from the measurement of $S_{21}$ and $S_{33}$ , respectively.
We also compute the participation of the bare modes by diagonalizing the linear part of the Hamiltonian \autoref{eq:H_bare_rwa} with the parameters of the multimode fit extracted from \subautoref{fig:fluxsweep}{b, c}.
The solid lines in \subautoref{fig:participation}{c, e} correspond to bare SAW modes, with the modes participating with less than \SI{2}{\percent}  shown in grey, and the black dashed line corresponds to $|u_{\mathrm{SQ},i}|^2$.
We find an excellent agreement with the measurements, down to $|u_{\mathrm{SQ},i}|^2 \simeq \SI{0.1}{\percent}$.
The hybridised mode $c_{12}$ is at the onset of multimode regime, with a maximum SQUID array resonator participation of \SI{20}{\percent}, and four bare SAW modes with participation above \SI{3}{\percent}.
The hybridised mode $c_{11}$ is deeper in the multimode regime. 
Indeed, it remains mainly acoustic all across the avoided crossing, with a maximum SQUID array resonator participation of only $|u_{\mathrm{SQ},i}|^2 \simeq \SI{4}{\percent}$.

\begin{figure*}
    \includegraphics[width = 1\linewidth ]{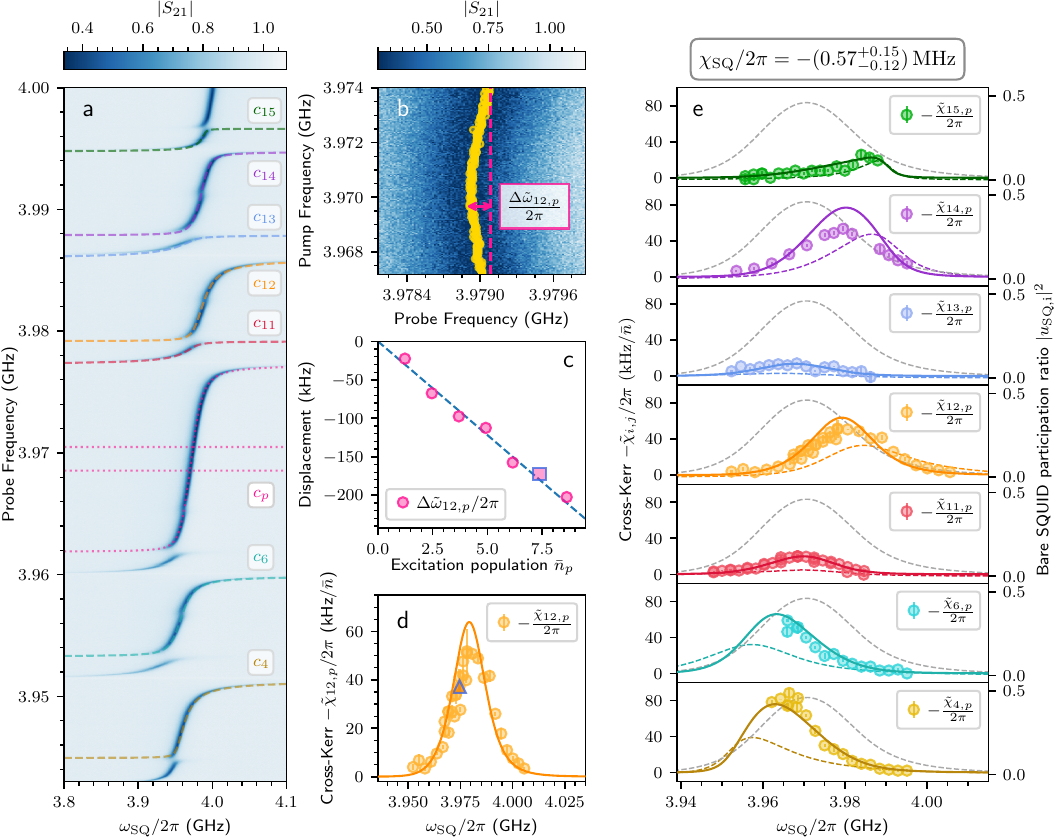}
    \caption{\label{fig:CK}
    \textbf{Cross-Kerr measurement versus bare SQUID array resonator frequency.}
    (a) Normalised transmission $S_{21}$ through the SQUID array resonator feedline, measured as a function of flux threading the SQUID array focusing on modes from $c_4$ to $c_{15}$.
    Dotted pink lines indicate the pumped mode $c_p \in [c_8 , \ c_{10}]$, while all the dashed lines represent all the probed modes.
    (b-d) Measurement protocol showing as an example $\tilde \chi_{12,p}$ the cross-Kerr between modes $c_{p}$ and $c_{12}$. 
    (b) We sweep the frequency across both the pumped mode ($c_p$) and the probed mode ($c_{12}$) to find the maximum frequency shift $\Delta f_{12,p}$.
    (c) We repeat this measurement for increasing pump power, and extract $\tilde \chi_{12,p}$ from a linear fit to the displacement against number of excitations in the pumped mode, $\bar{n}_p$.
    The square marker corresponds to the pump power used in (b).
    (d) We repeat the protocol outlined in (b,c) for different bare SQUID array resonator frequencies $\wsq$.
    The cross-Kerr value extracted in (c) is highlighted with a blue triangle marker.
    (e) Cross-Kerr as a function of $\wsq$, measured while pumping mode $c_p$, and probing seven different modes, using the same color coding as in panel (a).
    A global fit of all cross-Kerr data is performed (solid coloured lines), with the only free fitting parameter being the bare SQUID Kerr nonlinearity, yielding $\chi_\mathrm{SQ}/2\pi=0.57^{+0.15}_{-0.12}\, \SI{}{\mega\hertz}$. 
    The dashed lines in (e) show the bare SQUID participation ratio obtained from the linear model: the grey curve corresponds to the pumped mode, while the coloured curves correspond to the probed modes.
    Error bars on markers in (d) and (e) are obtained from the linear fit of displacements $\Delta \tilde \omega_{i,p}$.
    }
\end{figure*}

\textbf{Participation\textendash dissipation rate:}
We now focus on how the total dissipation rates $\tilde \kappa_j$ of the hybridised modes evolve as a function of the SQUID array resonator frequency $\omega_\mathrm{SQ}$.
As shown in \subautoref{fig:participation}{d, f}, an increase in dissipation rate is observed when the SQUID array resonator has the largest participation in the hybrid mode.
The total dissipation rates $\tilde \kappa_j$ of the hybrid mode $c_j$ can be approximated from the participation of the bare modes as
\begin{equation}
    \tilde \kappa_j \approx |u_{\mathrm{SQ},j}|^2 \, \kappa_\mathrm{SQ} + \sum_{i=1}^{29} |u_{i,j}|^2 \, \kappa_i \ ,
    \label{eq:hybrid_kappa}
\end{equation}
where $\kappa_{\mathrm{SQ}}$ is the total dissipation rate of the bare SQUID array resonator, and $\kappa_i$ denotes the dissipation rate of the $i$-th bare SAW mode.
As discussed in \autoref{subsec:appendix_deriv_dissip}, this expression neglects corrections arising from the overlap of the hybrid SAW mode profiles with the launcher IDT, which modify the effective external coupling~\cite{sletten2021quantum}.
The only unknown parameter in \autoref{eq:hybrid_kappa} is $\kappa_\mathrm{SQ}$, of which we do not have direct access when $\omega_\mathrm{SQ}$ is within the Bragg mirror bandwidth.
We perform a global fit of the dissipation rates $\tilde \kappa_j$ for all hybridised modes from $c_{10}$ to $c_{15}$ using \autoref{eq:hybrid_kappa}.
We fix all the participation ratios by diagonalising the linear part of Hamiltonian \autoref{eq:H_bare_rwa} with the parameters from the multimode fit of \subautoref{fig:fluxsweep}{d}.
All the bare SAW dissipation rates used in the estimation are reported in \autoref{fig:saw_dissip_appendix}.
We obtain 
$\kappa_\mathrm{SQ}/2\pi= \SI[separate-uncertainty = true]{8.8(7)}{\mega\hertz}$.

Measurements of SQUID array resonator participation and dissipation rates for four additional hybridised modes are presented in the Appendix (see \autoref{fig:deriv_dissip_appendix}), along with further details on the analysis procedure.
We note the presence of a few weakly coupled spurious modes, which may be attributed to higher-order transverse SAW modes.
These spurious modes are included in the extended analysis presented in \autoref{fig:double_peak_appendix}.

Remarkably, although the system operates in the weak coupling regime at the level of individual SAW modes, characterised by $g_i < \kappa_\mathrm{SQ}$, we observe hallmark signatures of strong coupling, such as vacuum Rabi splitting \cite{raimond2006exploring} (see \subautoref{fig:participation}{a}). 
This counterintuitive behavior arises because the dissipation of the SQUID array resonator is effectively distributed among the many mechanical modes to which it is simultaneously coupled. 
As a result, an \emph{effective strong coupling} emerges despite each mechanical mode being individually in the weak interaction regime with the SQUID array resonator. 

\subsection{Cross-Kerr Nonlinearities}

\textbf{Participation ratio and nonlinearity--theory:}
While the dissipation rate $\tilde \kappa_i$ of hybridised modes depends linearly on the participation ratio of the bare SQUID array resonator mode (\autoref{eq:hybrid_kappa}), the nonlinear interaction terms $\tilde \chi_{i,j}$ exhibit a quadratic dependence, as derived in \autoref{sec:bogoliubov_transformation}~\cite{nigg2012black}:
\begin{subequations}
\begin{gather}
    \tilde \chi_{i,i} = \chi_{\mathrm{SQ}} \, |u_{\mathrm{SQ},i}|^2 \, |u_{\mathrm{SQ},i}|^2\,, \label{eq:particip_KerrA} \\
    \tilde \chi_{i,j} = -2 \sqrt{\;\tilde \chi_{i,i}\;\tilde \chi_{j,j} } = 2 \chi_{\mathrm{SQ}} \, |u_{\mathrm{SQ},i}|^2 \, |u_{\mathrm{SQ},j}|^2 \, ,
    \label{eq:particip_Kerr}
\end{gather}
\end{subequations}
where $i\neq j$.

Already in the single-mode coupling regime, a large SQUID array resontor participation can result in a significant self-Kerr nonlinearity~\cite{yang2024mechanical}. 
However, achieving sizeable cross-Kerr terms requires substantial SQUID array resonator participation across multiple hybridised modes. 
Therefore, the multimode coupling regime is essential to simultaneously realise large cross-Kerr interactions at a fixed bare SQUID array resonator frequency~\cite{kuzmin2019superstrong}.

\textbf{Participation ratio and nonlinearity-measurements:}
Measurements of the cross-Kerr nonlinearity between different hybridised modes as a function of the bare SQUID array resonator frequency are presented in \autoref{fig:CK}.
To characterise the cross-Kerr interactions, we sweep a strong pump tone across the frequency range from hybridised mode $c_8 $ to $ c_{10} $ (denoted as $c_p$ in \subautoref{fig:CK}{a}), while weakly probing other eigenmodes to detect frequency shifts induced by the pump.
For each pump power, we perform two-tone spectroscopy by sweeping both the pump and probe frequencies, yielding a map of the mode response as shown in \subautoref{fig:CK}{b}. 
From this map, we extract the maximum negative frequency shift $\Delta \tilde{\omega}_{i,j}$ induced in the probed mode $ i $ due to the excitations in mode $ j $.
This protocol is repeated for increasing pump powers to extract the dependence of $\Delta \tilde{\omega}_{i,j}$ on pump power (see \subautoref{fig:CK}{c}).
The extraction of the cross-Kerr coefficient, defined as the frequency shift per excitation in the pumped mode, requires calibration of the input line attenuation to convert applied pump power to average excitation number in the mode.
We perform this calibration using a separate device \cite{peyruchat2025landau} connected via a microwave switch to the same measurement lines, following the procedure described in \autoref{sec:Attenuation_estimation}.
We repeat the full protocol across different bare SQUID array resonator frequencies to obtain the data markers shown in \subautoref{fig:CK}{d-e}.
The cross-Kerr measurement protocol is further detailed in \autoref{subsec:CK_meas_protocol} of the appendix.

We measure seven distinct cross-Kerr shifts as a function of $ \omega_{\mathrm{SQ}} $, as shown in \subautoref{fig:CK}{e}. 
The maximum cross-Kerr values of the probed modes range from 14 to \SI{88}{\kHz}. 
Notably, we observe non-zero cross-Kerr interactions for multiple SAW modes simultaneously (i.e. for a given bare SQUID array resonator frequency), which provides further evidence that the system operates in the multimode coupling regime.
To quantitatively analyse the data, we perform a global fit of all cross-Kerr measurements using~\autoref{eq:particip_Kerr}. 
The SQUID participation ratios are known from prior mode hybridisation analysis (\autoref{fig:fluxsweep}), and they are shown as dashed lines in \subautoref{fig:CK}{e}.
This leaves the bare SQUID Kerr nonlinearity  $\chi_{\mathrm{SQ}} $ as the only free parameter. 
The fit yields 
$\chi_\mathrm{SQ}/2\pi = 0.57^{+0.15}_{-0.12}\, \SI{}{\mega\hertz}$,
and the resulting curves (solid lines) show excellent agreement with the experimental data.

Additional self-Kerr measurements are provided in the appendix in \autoref{fig:self_kerr_appendix}.

\section{Discussion and Conclusion}

In this work, we have explored a nonlinear multimode mechanical system realised by coupling multiple confined SAW modes to a nonlinear, flux-tunable SQUID array resonator. We demonstrate that the device operates at the onset of the mechanical multimode (or superstrong) coupling regime.
By measuring the participation ratio of the SQUID array resonator in the hybridised modes, we show that both the induced mechanical nonlinearity and the total mode dissipation are accurately captured by this single metric. 
The ability to probe SAW and SQUID array resonator modes independently allows us to validate this model through direct measurements of both mode-dependent dissipation and cross-Kerr interactions between mechanical modes.

A key signature of the multimode coupling regime is that the SQUID array resonator participation ratio, $|u_{\mathrm{SQ},i}|^2$, remains well below 50\% even at maximum hybridisation with the mechanical modes.
In our device, in particular, we measure a maximum participation ratio as low as 4\%.
While this behaviour was previously modelled in \cite{moores2018cavity}, it had not been directly measured.
Here, we introduce a simple and general method to extract participation ratios from the derivatives of the hybrid mode frequencies with respect to the bare SQUID array resonator frequency. 
This approach is particularly relevant because participation-ratio-based modelling is a standard design tool in superconducting quantum circuits~\cite{minev2021energy}, yet experimental extraction is often challenging. 
Our method is broadly applicable, including to complex systems where numerical modelling is impractical.

Another distinctive feature of the multimode coupling regime is the simultaneous inheritance of nonlinearity by several mechanical modes. This is confirmed through two-tone cross-Kerr measurements, which reveal sizeable phonon-phonon interactions mediated by the nonlinear ancilla. 

Weakly nonlinear electromagnetic modes have already enabled a broad range of applications in quantum science, including parametric amplification~\cite{castellanos2008amplification, eichler2014controlling}, frequency conversion~\cite{chang2020observation}, quantum sensing~\cite{di2023critical, beaulieu2025criticality, cai2025quantum}, quantum simulation of Ising-like networks~\cite{puri2017quantum, alvarez2024biased}, non-reciprocal coupling schemes~\cite{hung2021quantum, slim2025programmable}, and even signature of quantum chaos~\cite{peyruchat2025landau}.
Our work extends these paradigms to the mechanical domain by moving beyond single-mode implementations of nonlinear mechanical systems~\cite{marti2024quantum}, and toward a regime where multiple massive modes interact nonlinearly in a controlled manner~\cite{andersson2022squeezing}.

\begin{figure}[] 
    \includegraphics[width = 1\linewidth ]{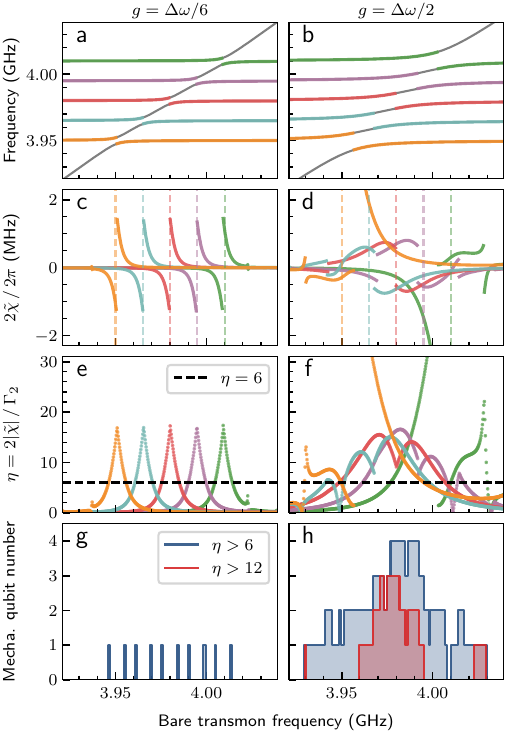}
    \caption{\label{fig:outlook}
    \textbf{Multiple mechanical qubits.}
    Numerical simulation of a multimode system analogous to the device presented in the work, with the SQUID array resonator replaced by a transmon qubit of self-Kerr $\chi_\mathrm{tr}/2\pi = -E_C/2h = \SI{-100}{\MHz}$.
    For simplicity, we consider only five SAW modes with uniform free spectral range $\Delta \omega /2\pi=\SI{15}{\MHz}$.
    Two coupling regimes are depicted: constant SAW-transmon coupling $g = \Delta \omega/6$ (left column) and $g = \Delta \omega/2$ (right column).
    (a,b) Fundamental transitions frequencies $\tilde \omega_{g,e} /2\pi$.
    For each bare transmon frequency, we identify the hybrid mode in which a given bare SAW mode has maximum participation, and color it according to the corresponding bare SAW mode.
    (c, d) Anharmonicity of the hybrid modes.
    The spectrum is computed by exact numerical diagonalisation within the two-excitation manifold.
    For each fundamental transition $\tilde \omega_{g,e}$, all corresponding transitions $\tilde \omega_{e,f}$ and their anharmonicities $2\tilde\chi = \tilde \omega_{e,f} - \tilde \omega_{g,e}$ are computed.
    Only the smallest absolute anharmonicity per mode is shown; the full spectrum is presented in the Appendix. 
    Bare SAW mode frequencies are indicated by vertical dashed lines.
    (e,f) Ratio of anharmonicity to decoherence rate $\eta = 2|\tilde \chi|/\Gamma_2$.
    We assume dissipation rates for bare SAW modes ($\kappa /2\pi = \Gamma_2 /\pi = \SI{50}{\kHz}$) and for the transmon $\kappa_{\rm tr} /2\pi = \Gamma_2 /\pi = \SI{300}{\kHz}$ \cite{sletten2019resolving}.
    Hybrid mode dissipation rates are calculated via \autoref{eq:hybrid_kappa}.
    A hybrid mode is designated as a qubit if $\eta >6$ (above the black dashed line) \cite{yang2024mechanical}.
    (g,h) Number of mechanical qubits as a function of bare transmon frequency, with transmon participation ratio $|u_{\rm{tr}}|^2$ below $20 \%$.
    Blue indicates $ \eta > 6 $, while red indicates $ \eta > 12$.
    }
\end{figure}

\textbf{Towards multiple mechanical qubits:}
Strong single-mode mechanical nonlinearity has recently been demonstrated using a bulk acoustic resonator coupled to a transmon qubit~\cite{yang2024mechanical}. Here, we explore the feasibility of extending our current platform to enable multiple mechanical qubits by replacing the weakly nonlinear SQUID array resonator with a single SQUID of reduced Josephson energy $E_{\rm J}$. This yields a transmon qubit with excitation frequency around \SI{4}{\GHz} and anharmonicity $-2\chi_\mathrm{tr} \approx E_C/\hbar \approx 2\pi \cdot \SI{200}{\MHz}$. Notably, the capacitance of the SQUID array resonator’s IDT in our current design—approximately \SI{70}{\fF}—is already within the typical range for transmon qubits.

However, replacing the weakly nonlinear SQUID array resonator with a transmon pushes the system into a strongly nonlinear regime where our previously used participation-ratio-based analysis becomes inapplicable.
This approach treats the nonlinearity as a perturbation~\cite{nigg2012black, kuzmin2019superstrong, minev2021energy}, which breaks down when $|\chi_\mathrm{tr}| \gtrsim g$~\cite{yilmaz2024energy}.
Indeed, while \autoref{eq:particip_Kerr} predicts that the hybrid modes self- and cross-Kerr always retain the sign of the bare nonlinearity, this no longer holds in the strongly nonlinear case, where a straddling regime may emerge, inducing sign changes in the nonlinearities~\cite{koch2007charge, boissonneault2010improved}.

To address these challenges, we model the transmon and each SAW mode as three-level systems and employ exact numerical diagonalisation to compute the low-energy excitation spectrum.
This enables us to identify parameter regimes where multiple mechanical modes acquire sufficient anharmonicity and retain enough coherence to function as qubits.

For simplicity, we simulate five SAW modes with uniform free spectral range (FSR) $\Delta\omega$ and equal coupling $g$.
We consider two coupling scenarios (illustrated in \autoref{fig:outlook}):
(i) a weak-coupling regime where the SAW-transmon coupling $g$ is much smaller than the free spectral range $\Delta\omega$ of the SAW cavity, $g = \Delta \omega/6$ (left column), and
(ii) a multimode coupling regime where $g = \Delta \omega/2$ (right column).

In the single-mode coupling case, only one mechanical mode strongly hybridises with the transmon at a time, resulting in well-separated anharmonicity peaks for each hybrid mode, with a maximum value of $2\tilde\chi = \pm g(2 - \sqrt{2})$~\cite{yang2024mechanical} (\subautoref{fig:outlook}{c}). The hybrid-mode decoherence rate is estimated as $\Gamma_2 = \tilde{\kappa}/2$, using \autoref{eq:hybrid_kappa} and assuming a typical transmon dissipation rate $\kappa_{\rm tr}/2\pi = \SI{300}{\kHz}$~\cite{sletten2019resolving} (\subautoref{fig:outlook}{e}).

We define a hybrid mode to be qubit-like if its anharmonicity-to-decoherence ratio $\eta = 2\tilde{\chi}/\Gamma_2$ exceeds~6~\cite{yang2024mechanical}, and further restrict to predominantly mechanical modes by requiring transmon participation $|u_\mathrm{tr}|^2 < 20\%$. In this regime, only one such mechanical qubit can be realised at a time (\subautoref{fig:outlook}{g}).

In contrast, in the multimode coupling case, all SAW modes hybridise simultaneously with the transmon, leading to a more complex anharmonicity structure (\subautoref{fig:outlook}{d}). Although the coupling increases, the maximum achievable anharmonicity remains comparable to the weak-coupling case. Crucially, however, all modes acquire significant anharmonicity. 
Accounting for hybrid-mode dissipation, we find that placing the transmon frequency at the center of the SAW cavity bandwidth enables all five SAW modes to cross the qubit threshold (\subautoref{fig:outlook}{f}). Up to four of these modes satisfy the condition $|u_\mathrm{tr}|^2 < 20\%$, thus qualifying as predominantly mechanical qubits (\subautoref{fig:outlook}{h}).

Upgrading our current platform to realise multiple mechanical qubits is within realistic experimental reach.
The required increase in FSR and coupling—approximately a factor of two—can be achieved by halving the effective SAW cavity length.
In addition, although we simulated only five modes with $\Delta\omega = \SI{15}{\MHz}$, more than ten longitudinal SAW modes can be confined within the bandwidth used in this work.

In our current design, the dissipation of the SQUID array resonator is dominated by coupling to the readout feedline (see \autoref{fig:SQ_dissip_appendix}). 
This can be mitigated by reducing the coupling capacitance.
Moreover, with improved fabrication techniques—such as surface cleaning prior to qubit deposition~\cite{kono2024mechanically}—the internal loss rate of the transmon could be reduced down to $\kappa_\mathrm{tr}/2\pi=\SI{300}{\kHz}$~\cite{sletten2019resolving}, as assumed in our simulations.
Finally, qubit dissipation may be decreased even further by employing flip-chip architectures~\cite{satzinger2018quantum}, which further suppress dielectric and radiation losses~\autoref{fig:appendix_qubits_flipchip}.

Altogether, leveraging the multimode coupling regime in SAW-based systems offers a compelling route toward realizing acoustic qubits with always-on self- and cross-Kerr interactions and long-range connectivity. This platform is particularly attractive for quantum simulation of quantum fluids of light~\cite{carusotto2013quantum}. 
Mechanical qubits confined to the surface provide further advantages over bulk implementations, including improved accessibility, the ability to support multiple coupling points~\cite{andersson2019non}, and tighter confinement of acoustic mode volumes. These features enhance prospects for strong coupling to elusive systems, such as charge or spin degrees of freedom~\cite{schutz2017universal}, and enable integration with phononic waveguides for scaling~\cite{wang2024perspectives}.

\subsection*{Acknowledgments}
We thank Orjan Ameye, Santiago Salazar Jaramillo, Oded Zilberberg and Gustav Andersson for stimulating discussion regarding this work.
P.S. acknowledges support from the Swiss National Science Foundation (SNSF) through the project UeM019-16 – 215928, and through the grants Ref. No. 200021\_200418 and Ref. No. 206021\_205335, and from the Swiss State Secretariat for Education, Research and Innovation (SERI) under contract number 01042765 SEFRI MB22.00081.
This work was supported by the Knut and Alice Wallenberg foundation through the Wallenberg Center for Quantum Technology (WACQT), and the Swedish Research Council, VR. The device was fabricated at MyFab Chalmers.

\subsection*{Author Contributions}
M.S., L.P., P.D. and P.S. designed the experiment. 
M. S. designed and fabricated the devices. 
M.S., C.B. and V.J. performed preliminary device measurements. 
L.P. and  R.M.M. performed the measurements presented in the manuscript. 
L.P., R.M.M. and M.S. analysed the data.
L.P. and R.M.M. developed the theoretical models.
P.S. and P.D. supervised the project. 
L.P., M.S., R.M.M. and P.S. wrote the manuscript with input from all authors.

\appendix

\section*{Appendix}

\section{System parameters and variables}

In \autoref{tab:table_symbols} the symbols employed in this work are summarised. In addition, \autoref{tab:geom_device_params} reports the geometric and cooldown-independent parameters of the device, while \autoref{tab:Device_setup_parameters} shows the parameters extracted in the characterisation of the device and of the measurement setup throughout several cooldowns.
Finally, in \autoref{tab:Figs_CD} each figure is matched with the cool downs when the data displayed were acquired. 

\begin{table}[t]
\centering
\begin{ruledtabular}
\begin{tabular}{cp{6cm}c}
\textbf{Symbol} & \textbf{Description} & \textbf{Unit} \\
\hline 
\\[0.5pt]
\( \phi_X \) & Magnetic flux threading SQUID array loops & \SI{}{\weber} \\
\( \omega_{SQ}/2\pi \) & Bare SQUID frequency & \SI{}{\hertz} \\
\( a; a^\dagger \) & SQUID array resonator annihilation and creation operators &  \\
\( \kappa_{\rm SQ}/2\pi \) & Total dissipation rate of the bare SQUID array resonator & \SI{}{\hertz} \\
\( \kappa_{\rm SQ,ext}/2\pi \) & External dissipation rate of the bare SQUID array resonator & \SI{}{\hertz} \\
\( \kappa_{\rm SQ,int}/2\pi \) & Internal dissipation rate of the bare SQUID array resonator & \SI{}{\hertz} \\
\( \chi_{SQ}/2\pi \) & SQUID array resonator self-Kerr & \SI{}{\hertz} \\
\\[1pt]
\( \omega_i/2\pi \) & Bare SAW frequencies & \SI{}{\hertz} \\
\( \Delta\omega_i/2\pi \) & Free spectral range of bare SAW modes & \SI{}{\hertz} \\
\( b_i; b_i^\dagger \) & SAW mode annihilation and creation operators &  \\
\( \kappa/2\pi \) & SAW mode total dissipation rate& \SI{}{\hertz} \\
\( \kappa_{\text{ext}}/2\pi \) & SAW mode external dissipation rate & \SI{}{\hertz} \\
\( \kappa_{\text{int}}/2\pi \) & SAW mode internal dissipation rate& \SI{}{\hertz} \\
\( g_i/2\pi \) & Coupling strength between the electromagnetic and mechanical modes & \SI{}{\hertz} \\
\\[1pt]
\( |u_{\text{SQ},j}|^2 \) & Participation of SQUID array resonator in the \( j \)-th hybrid modes & \\
\( |u_{i,j}|^2 \) & Participation of \( i \)-th SAW mode in \( j \)-th hybrid mode &  \\
\( \tilde{\omega}_i/2\pi \) & Frequency of hybrid modes & \SI{}{\hertz} \\
\( c_i; c_i^\dagger/2\pi \) & Hybrid mode annihilation and creation operators &  \\
\( \tilde{\kappa}_j/2\pi \) & Total dissipation rate of the hybrid modes & \SI{}{\hertz} \\
\( \tilde{\chi}_{ij}/2\pi \) & Kerr-coefficients among the hybrid modes & \SI{}{\hertz} \\
\( \Delta \tilde{\omega}_{i,j}/2\pi \) & Maximum displacement of \( i \)-th hybrid mode when pumping the \( j \)-th hybrid mode & \SI{}{\hertz} \\
\\[1pt]
\( {\kappa}_{\rm tr}/2\pi \) & Total dissipation of transmon qubit & \SI{}{\hertz} \\
\( |u_{\text{tr}}|^2 \) & Participation of trasmon qubit in the hybrid modes & \\
\( {\chi}_{\rm tr}/2\pi \) & Transmon qubit self-Kerr  & \SI{}{\hertz} \\
$\eta = 2\tilde \chi / \Gamma_2$ & Anharmonicity to decoherence ratio  &  \\

\\[1pt]
\( S_{21} \) & Normalised transmission notch-type coefficient measured from the SQUID array resonator feedline &  \\
\( S_{33} \) & Normalised reflection coefficient measured from launcher IDT feedline &  \\
\end{tabular}
\end{ruledtabular}
\caption{\textbf{Table of symbols.}}
\label{tab:table_symbols}
\end{table}

\begin{table}[t]
\centering
\begin{ruledtabular}
\begin{tabular}{l p{4cm} c c l}
\textbf{Parameter} & \textbf{Description} & \textbf{Value} & \textbf{Unit} \\
\hline
\\[0.1pt]
$\lambda_0$ & Wavelength of the central SAW mode & $\SI{720}{}$ & \SI{}{\nm} \\
$\omega^c_0/2\pi$ & Central frequency of the SAW cavity & \SI{3.967(9)}{} & \SI{}{\giga\hertz} \\
$r$ & Single finger reflectivity & $\SI{3.9}{\percent}$ &  \\
$\Delta_{m}/2\pi$ & Bragg mirror bandwidth &  \SI{99}{} & MHz \\
$d$ & Distance between Bragg mirrors & \SI{155.4}{} & \SI{}{\micro\meter} \\
$L_c$ & Effective cavity length & \SI{165}{} & \SI{}{\micro\meter} \\
$N_m$ & Number of fingers in each Bragg mirror &  $500$  &  \\
$N_p$ & Number of pairs of split-fingers in launcher IDT & 30 &  \\
$N_p$ & Number of pairs of split-fingers in coupler IDT & 14 &  \\
$N_s$ & Number of SQUIDs in the array & $12$ & \\
$K^2$ & Effective electromechanical coupling in GaAs & $\SI{0.07}{\percent}$ &  \\
\end{tabular}
\end{ruledtabular}
\caption{\textbf{Table of geometric and cooldown-independent parameters.}}
\label{tab:geom_device_params}
\end{table}

\begin{table*}[t]
\centering
\begin{ruledtabular}
\begin{tabular}{p{2.0cm} @{\hskip 0.5cm} p{5.0cm} @{\hskip 0.5cm} c c c c l}
\textbf{Parameter} & \textbf{Description} & \textbf{20240913CD} & \textbf{20240927CD} & \textbf{20241206CD} & \textbf{20250214CD} & \textbf{Unit} \\
\hline
\\[0.1pt]
$\omega_{\mathrm{SQ},0}/2\pi$ & SQUID array resonator frequency at sweetspot & $4.70 \pm 0.04$ & $4.709 \pm 0.003$ & $4.7095 \pm 0.0002$ & / & GHz \\
\\[0.01pt]
$\alpha_{\mathrm{SQ}}$   & Voltage-to-reduced-flux conversion factor        & $1.14 \pm 0.04$ & $1.083 \pm 0.003$ & $1.0852 \pm 0.0008$ & / & rad/V \\
\\[0.01pt]
$V_{\mathrm{SQ},0}$      & Voltage offset at sweetspot       & $2.6 \pm 3$     & $35 \pm 2$        & $39.5 \pm 0.2$       & / & mV \\
\hline
\\[0.1pt]
$\kappa_{\mathrm{SQ}} / 2\pi$ & Bare SQUID dissipation inside Bragg bandwidth & $8.8 \pm 0.7$ & / & / & / &  \SI{}{\mega\hertz} \\
\\[0.01pt]
$\chi_{\mathrm{SQ}}/ 2\pi$    & Bare SQUID array resonator nonlinearity & / & / & $0.57^{+0.15}_{-0.12}$    & / &  \SI{}{\mega\hertz} \\
\hline
\\[0.01pt]
$A^{\rm SQ}$ & Attenuation of the SQUID array resonator feedline (port 1) & $-78.8 \pm 2$   & $-97 \pm 1.5$ & $-101 \pm 1$ & $-102 \pm 1.5$ & dB \\
$A^{\rm IDT}$ & Attenuation of the launcher IDT feedline (port 3) & $-78.8 \pm 2$  & $-97 \pm 2$ & / & $-103.6 \pm 2$ & dB \\
\end{tabular}
\end{ruledtabular}
\caption{\textbf{Device and measurement setup parameters.} The measurements in this work were performed over several cooldowns.
This table reports the parameters of the device and the attenuation of the input feedlines employed in the different cooldowns (CDs).
"/" denotes missing values for the specific cooldown.
In the 20240927CD and subsequent cooldowns, we added 20dB attenuation at room temperature compared to the 20240913CD.
Other changes and calibration methods are explained in \autoref{sec:Attenuation_estimation}.
}
\label{tab:Device_setup_parameters}
\end{table*}

{
\renewcommand{\arraystretch}{1.15}
\begin{table}[t]
\centering
\begin{ruledtabular}
\begin{tabular}{l p{4.3cm} c }
\textbf{Figure} & \textbf{Description} & \textbf{CD} \\
\hline
\multicolumn{3}{l}{\textit{Main text}} \\
Fig. 2 & \textbf{Bare SAW modes} & 20240913CD  \\
Fig. 3 a) & \textbf{Flux sweep of the SQUID array resonator across SAW modes} & 20240927CD  \\
Fig. 3 b), c) & \textbf{Flux sweep of the SQUID array resonator across SAW modes} & 20240913CD  \\
Fig. 4 & \textbf{Characterisation of the multimode coupling regime} & 20240913CD  \\
Fig. 5 & \textbf{Cross-Kerr measurement versus bare SQUID array resonator frequency}& 20241206CD  \\
\\[0.1pt]
\hline
\multicolumn{3}{l}{\textit{Appendix}} \\
\autoref{fig:attenuation_calib} & \textbf{Input line attenuation calibration} & 20241206CD  \\
\autoref{fig:bare_SAW_dissip_appendix} & \textbf{Frequency and dissipation rates of bare SAW modes} & 20240913CD  \\
\autoref{fig:saw_fsr_appendix} & \textbf{Frequency difference between SAW modes} & 20240913CD  \\
\autoref{fig:saw_dissip_power_appendix} & \textbf{Internal dissipation rate of SAW modes as a function of input power} & 20250214CD  \\
\autoref{fig:SQ_dissip_appendix} & \textbf{Dissipation of SQUID Array versus Flux} & 20240927CD  \\
\autoref{fig:deriv_dissip_appendix} & \textbf{Additional data for Fig. 4} & 20240913CD  \\
\autoref{fig:double_peak_appendix} & \textbf{Mode participation and dissipation rate analysis in presence of a spurious SAW mode} & 20240913CD  \\
\autoref{fig:self_kerr_appendix} & \textbf{Self-Kerr measurement at one flux value} & 20241206CD  \\
\autoref{fig:cross_kerr_appendix} & \textbf{Cross-Kerr measurement at one flux value} & 20241206CD  \\

\end{tabular}
\end{ruledtabular}
\caption{
\textbf{
Association between measurement and cooldown.
}}
\label{tab:Figs_CD}
\end{table}
}

\section{Theory}

\subsection{Model}

\subsubsection*{Hamiltonian, Bogoliubov, nonlinearity expansion}
\label{sec:bogoliubov_transformation}

The Hamiltonian of a linear bosonic mode, $a_0$, coupled to $n$ other uncoupled linear bosonic modes, $a_{i>0}$, can be written as
\begin{equation}
    H_0/\hbar = \wsq \creat{a_0} \annih{a_0} 
    + \sum_{i>0}^{n} \omega_i \creat{a_i} \annih{a_i} - (\annih{a_0} - \creat{a_0}) \sum_{i>0}^n g_i (\annih{a_i} - \creat{a_i}) \,,
\label{Eq:Full_Linear_Hamiltonian}
\end{equation}
with the usual bosonic commutation relations $[\annih{a_k},\creat{a_l}]=\delta_{kl}$.
To map the model to our device we can identify $\annih{a_0}$ with operator of SQUID array resonator, and $\annih{a_{i>0}}$ with the ones for the $i-th$ bare SAW mode.
This Hamiltonian takes into account particle non-conserving terms such as $\annih{a_0}\annih{a_i}$ and $\creat{a_0}\creat{a_i}$, so a priori the diagonalisation should be performed via Bogoliubov transformation.
This protocol is usually employed in Many-Body Condensed Matter Physics to diagonalise quadratic Hamiltonians by combining the original creation and annihilation operators via a symplectic transformation $Q$, preserving the original bosonic algebra. 

In the first place, the Hamiltonian \autoref{Eq:Full_Linear_Hamiltonian} can be rewritten as a quadratic form,
\begin{equation}
    H_0/\hbar = \frac{1}{2} 
    \begin{pmatrix}
    \creat{\boldsymbol{a}} & \annih{\boldsymbol{a}} \\
    \end{pmatrix}
    \begin{pmatrix}
    T & U \\
    U^* & T^*
    \end{pmatrix}
    \begin{pmatrix}
    \annih{\boldsymbol{a}} \\
    \creat{\boldsymbol{a}} \\
    \end{pmatrix}
    + \rm const .
\label{Eq:General_hamiltonian_matrix_form}
\end{equation}
Here "$\rm const.$" is a diagonal matrix with the vacuum energy of the different fields, and we define $\left(\creat{\boldsymbol{a}} \ \annih{\boldsymbol{a}}\right) = \left( \creat{a_0},\creat{a_1},\cdots,\creat{a_{n}},\ \annih{a_0},\annih{a_1}, \cdots,\annih{a_{n}} \right)$, the hermitian adjoint of this vector as  $\begin{pmatrix}
    \annih{\boldsymbol{a}} \\
    \creat{\boldsymbol{a}} \\
    \end{pmatrix} = \creat{\left(\creat{\boldsymbol{a}} \ \annih{\boldsymbol{a}}\right)} $, and the matrices $T$ and $U$ in the basis $\mathcal{B}_{a} = \{ \annih{a_0},\annih{a_1}, \cdots,\annih{a_{n}} \}$ can be written as
\begin{equation}
    T_{\mathcal{B}_{a}} = \begin{pmatrix}
\wsq & g_1 & \cdots & g_{n} \\
g_1 & \omega_1 & \cdots & 0 \\
\vdots & \vdots & \ddots & \vdots \\
g_{n} & 0 & \cdots & \omega_{n} \\
\end{pmatrix}
\label{Eq:T_and_U_matrices}
\end{equation}
\begin{equation}
    \text{and} \
    U_{\mathcal{B}_{a}} =  \begin{pmatrix}
0 & -g_1 & \cdots & -g_{n} \\
-g_1 & 0 & \cdots & 0 \\
\vdots & \vdots & \ddots & \vdots \\
-g_{n} & 0 & \cdots & 0 \\ 
\end{pmatrix} .
\end{equation}
Following the protocol introduced in \cite{del2004quantum}, we write the commutator of the $\annih{a}$-bosons in a matrix form and obtain the metric $\Omega$, which guarantees that the transformation $Q$ preserves the commutation relations
\begin{equation}
\begin{aligned}
    [X, \creat{X}] \equiv X\creat{X}- (\creat{X})^T X^T = \Omega   ,
\end{aligned}    
\end{equation}
\begin{equation}
    \Omega =  
    \begin{bmatrix}
     \mathbb{I}_{n+1} & 0 \\
     0 & -\mathbb{I}_{n+1}
    \end{bmatrix},
\end{equation}

where $X \equiv \left(\creat{\boldsymbol{a}} \ \annih{\boldsymbol{a}}\right)$, for compactness of the notation.
 Therefore, the orthonormalisation condition for the transformation $Q$, (also referred to as symplectc condition) reads as 
\begin{equation}
    Q,\Omega \in \mathbb{M}_{2(n+1)\times2(n+1)}( \mathbb{C}) \quad :   \
    Q \Omega \creat{Q} = \Omega, \quad
\label{sympl_condition_1}   
\end{equation}

This condition can be recast into the following relation on the inverse of the transformation
\begin{equation}
    Q^{-1} = \Omega^{-1} \creat{Q} \Omega \ \iff \ \creat{Q} = \Omega^{-1} Q^{-1} \Omega  .
    \label{sympl_condition_2}
\end{equation}

Leveraging \autoref{sympl_condition_2}, the diagonalisation of the hamiltonian matrix $\text{H}$ can be rewritten as 

\begin{equation}
\begin{aligned}
    \creat{Q}\text{H}Q=D &\rightarrow  \ (\Omega^{-1} Q^{-1} \Omega) \text{H}Q=D & \\
    &\rightarrow \ Q^{-1} (\Omega \text{H}) Q= \Omega D,
\end{aligned}
\end{equation}
where $D$ is the diagonal matrix with the eigenvalues of $\Omega \text{H}$. In the most general context $Q$ can be obtained from $Q_{\Omega \text{H}}$, the transformation diagonalising $\Omega \text{H}$, by means of subsequent rotations within the eigenspaces associated to each distinct eigenvalue. However, thanks to the symmetries of the model, $\Omega \text{H}$ is \textbf{not singular and its spectrum is not degenerate}. Therefore, $Q$ coincides with $Q_{\Omega \text{H}}$, provided that the column eigenvectors of $\Omega \text{H}$ are ortho-normalisised (w.r.to the conventional metric, $\mathbb{I}_{2(n+1)\times2(n+1)}$). 
In addition, the structure of $Q$ can be further constrained in the block form
\begin{equation}
    Q =\begin{pmatrix}
    u & v \\
    v^* & u^*
    \end{pmatrix} \ \text{with} \ u, v \in M_{n\times n}( \mathbb{C}),
\end{equation}
and by restricting the coefficients to be real, the transformation connecting the \textbf{a} and the \textbf{c} bosons reads as
\begin{equation}
    \begin{pmatrix}
    \annih{\boldsymbol{a}} \\
    \creat{\boldsymbol{a}} \\
    \end{pmatrix}
    =
    \begin{pmatrix}
    u & v \\
    -v & u
    \end{pmatrix}
    \begin{pmatrix}
    \annih{\boldsymbol{c}} \\
    \creat{\boldsymbol{c}} 
    \end{pmatrix} \text{with} \ u, v \in M_{n\times n}( \mathbb{R}),
\label{Eq:transformation_a_to_c_bosons}
\end{equation}

where $\left(\annih{\boldsymbol{c}} \ \creat{\boldsymbol{c}}\right) = \left(\annih{c_0},\annih{c_1}, \dots,\annih{c_{n}}, \creat{c_0},\creat{c_1},\dots,\creat{c_{n}}\right)$.

\textbf{Kerr term.} Now we take into account that the common mode $\annih{a_0}$ is, in fact, nonlinear. Namely, we perturbatively transform the self-Kerr term $\chi_{\rm SQ}/6 \left(\annih{a_0} + \creat{a_0}\right)^4$ according to \autoref{Eq:transformation_a_to_c_bosons}.
By selecting the terms of the form $(\creat{c_i}\annih{c_i})^2$ and $\creat{c_i}\annih{c_i}\creat{c_j}\annih{c_j}$, we find an expression for the self- and cross-Kerr of the $\annih{\boldsymbol{c}}$-bosons on participation of $\annih{a_0}$ and $\creat{a_0}$ in the hybrid modes, 
\begin{subequations}
\begin{gather}
        \frac{\chi_{\rm SQ}}{6}(\annih{a_0}+\creat{a_0})^4 \rightarrow \chi_{ii} (\creat{c_i}\annih{c_i})^2 + \chi_{ij} (\creat{c_i}\annih{c_i}\creat{c_j}\annih{c_j}) + \dots
        \label{quartic_term_expansion} , \\ 
        \chi_{ii} = \chi_{\rm SQ} ( u_{0,i} - v_{0,i})^4 \ , \\ 
        \chi_{ij} = 2 \chi_{\rm SQ} (u_{0,i} - v_{0,i})^2(u_{0,j} - v_{0,j})^2 ,
        \label{Eq:self_cross_kerr}
\end{gather}
\end{subequations}
where $\chi_{ii}$ and $\chi_{ij}$ are the self- and the cross-Kerr respectively. The cross-Kerr term complies with the relation provided in \cite{nigg2012black}.

\textbf{Simplified model.} Following the scheme introduced above to numerically diagonalise the Hamiltonian matrix $\text{H}$, for the coupling regime of our device, we find that the coefficients of the $v$ matrix can be neglected with respect to the ones of the $u$ matrix. In fact, the model \eqref{Eq:Full_Linear_Hamiltonian} reduces to a particle conserving Hamiltonian, equivalent to performing rotating-wave approximation
\begin{equation}
        H_0/\hbar \approx \wsq \creat{a_0} \annih{a_0} 
    + \sum_{i>0}^{n} \omega_i \creat{a_i} \annih{a_i} + \sum_{i>0}^n g_i (\creat{a_0}\annih{a_i} + \creat{a_i}\annih{a_0}),
\label{Eq:Full_Linear_Hamiltonian_RWA}
\end{equation}
which in matrix form simplifies to 
\begin{equation}
    H/\hbar \approx \creat{\boldsymbol{a}}T_{\mathcal{B}_a} \annih{\boldsymbol{a}} + \mathrm{const},
\label{Eq:RWA_hamiltonian_matrix}
\end{equation}
with the same meaning of the symbols as above.
Thus, the Hamiltonian \autoref{Eq:RWA_hamiltonian_matrix} is diagonalised just by the $u$ transformation.
This model is equivalent to  \autoref{eq:H_bare_rwa} after identifying $a_0 \equiv a \, \text{, and} \, a_i \equiv b_i \, \text{for } \, i>0 $.
Finally, self- and cross-Kerr relations can be simplified as
\begin{subequations}
\begin{gather}
    \chi_{ii} \approx \chi_{SQ} \, (|u_{0,i}|^2)^2, \ \\ 
    \chi_{ij} \approx 2 \, \chi_{SQ}  \,|u_{0,i}|^2 \,|u_{0,j}|^2 .
    \label{Eq:self_cross_kerr_approx}
\end{gather}
\end{subequations}

\subsubsection*{Participation from derivative}
The participation of the SQUID resonator in the hybrid modes can be extracted from the derivative of the corresponding eigenenergy with respect to the bare SQUID frequency $\omega_{\text{SQ}}$.
First, we restrict to the single excitation subspace, and consider eigenstates $\ket{\tilde{\psi_i}}$ with eigenenergies $\tilde{\omega_{i}}$. Then, by leveraging the Hellmann-Feynman theorem~\cite{politzer2018hellmann} and the fact that the projector onto a space of finite size is a bounded operator, thus continuous, we can write as follows:
\begin{equation}
\begin{aligned}
     \derivative{\tilde{\omega}_i}{\wsq} & = \derivative{}{\wsq} \bra{\tilde{\psi_i}}{H_0/\hbar}\ket{\tilde{\psi_i}} 
    = \bra{\tilde{\psi_i}} \derivative{H_0/\hbar}{\wsq} \ket{\tilde{\psi_i}} \\ & = \bra{\tilde{\psi_i}}{n}_{0}\ket{\tilde{\psi_i}} = u_{0,i}^*u_{0,i} \, ,
\end{aligned}
\label{Eq:Participation_from_derivative}
\end{equation}
where $n_{ 0}$ is the number operator of mode $a_0$, associated to the SQUID array resonator.

\subsubsection*{$\rm{SQUID}$ array resonator characteristics}
\begin{equation}
\begin{aligned}
    \wsq(\phi_X)/2\pi & = \omega_\mathrm{SQ,0} \sqrt{|\cos((\phi_X -\phi_0)/\Phi_0)|} /2\pi \\ &= \omega_\mathrm{SQ,0} \sqrt{|\cos(\alpha(V-V_0))|} /2\pi 
\end{aligned}
\label{eq:SQUID_characterisation}
\end{equation}
where $\omega_{\mathrm{SQ}, 0}$ is the maximum frequency of the SQUID array resonator at zero flux, $\phi_X$ is the magnetic flux threading the SQUID loops, $\phi_0$ the residual flux offset, and $\Phi_0$ the magnetic flux quantum. 
Experimentally, the magnetic flux bias is applied by delivering a DC voltage across a $\SI{\approx1}{\kohm}$ bias resistor in series with an external superconducting coil located beneath the sample.
As a result, the relevant calibration parameters for the SQUID array resonator are the conversion factor from voltage to reduced flux, $\alpha$, and the voltage offset corresponding to the zero-flux point, $V_0$.

\subsubsection*{Input Output relations}

The input-output relation employed to characterise the transmission signal in notch configuration, $S_{21}$, measured through port 1 and 2, is~\cite{chen2022scattering}
\begin{equation}
    \underset{\omega_r, \phi, \kappa_{\rm tot}, \kappa_{\rm c}}{S_{21}(\omega)}= 1 - \frac{\kappa_{\rm c} }{\left(\kappa_{\text{tot}} + 2i(\omega - \omega_{\rm r})\right)}\times\frac{e^{i\phi}}{\cos(\phi)},
\label{Eq:Notch_coefficient_single_mode_FANO}
\end{equation}
where $\omega$ is the frequency of the probe tone, $\omega_{\rm r}$ is the frequency of the mode, $\kappa_{\rm tot}$ is the total dissipation of the mode, $\kappa_{\rm c}$ is the external dissipation rate of the mode to the microwave line connecting ports 1 and 2.
Finally, $\phi$ is the phase difference between the main signal and the background signals that induce Fano interference in the measured signal \cite{probst2015efficient, rieger2023fano}.

The input-output relation employed to characterise the reflection signal measured from port 3 is
\begin{equation}
    \underset{\omega_r, \phi, \kappa_{\rm tot}, \kappa_{\rm c}}{S_{33}(\omega)}= \frac{\left(\kappa_{\text{tot} } - \kappa_{c }\, (1+ e^{i \phi})\right)+2i\left(\omega-\omega_{\rm r}\right)}{\left(\left(\kappa_{c }+\kappa_{\text{i} }\right)+2 i\left(\omega-\omega_{\rm r}\right) \right)\cos(\phi)} \ ,
\label{Eq:reflection_coeff_single_mode}
\end{equation}
where here $\kappa_{\rm c}$ is the coupling rate of the mode to the feed line connected to port 3.

\subsubsection*{Conversion to excitation number}
The relation to convert the input power into the steady-state average photon/phonon population of a linearly driven resonator is
\begin{equation}
    \bar{n}=\frac{1}{\hbar \omega_r} \frac{\kappa_{\text{c}}}{\Delta^2+\left(\kappa_{\text{tot}}/2\right)^2} \times P_{\text {in,W }} \times 10^{-|\rm A|/10} ,
\label{Eq:Photon_number_conversion}
\end{equation}
where $\Delta=\omega-\omega_r$ is the detuning of the probe tone, $\omega$, from the resonator frequency $\omega_r$, $P_{\rm in,W}$ the power of the probe tone at room temperature in Watt, and $\rm A$ is the total attenuation of the line in \SI{}{\dB} as calibrated with the protocol described below in section \autoref{sec:Attenuation_estimation}.

\subsection{Outlook: mechanical qubits}

\begin{figure}[h]
    \includegraphics[width = 1\linewidth ]{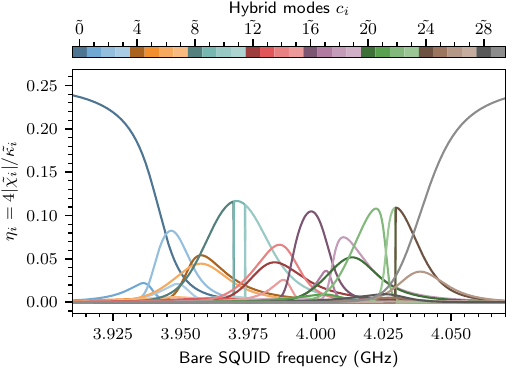}
    \caption{\label{fig:appendix_qubit_weak_chi}
    \textbf{Ratio between anharmonicity $2|\tilde \chi_i|$ and decoherence rate $\tilde \kappa_i/2$ for all hybrid modes of the SQUID array device presented in this work.}
    We compute the anharmonicity to decoherence rate ratio $\eta = 4|\tilde \chi_i | / \tilde\kappa_i $ using \autoref{eq:hybrid_kappa} and \autoref{eq:particip_KerrA}.
    Different colors indicate the 30 hybrid modes $c_i$.
    }
\end{figure}

\textbf{From weakly nonlinear modes to mechanical qubits.}
We have demonstrated that our platform offers control on acoustic mode nonlinearities by controlling the SQUID array resonator frequency.
The effective nonlinearity of the hybridised modes is directly determined by the nonlinearity of the bare SQUID resonator and by the \emph{square} of its participation ratio in each hybridised mode (\autoref{eq:particip_KerrA}).
To facilitate comparison with the literature on qubit systems~\cite{blais2021circuit, yang2024mechanical}, we define a dimensionless parameter of the quantum character of a mode as $\eta = 2 |\chi|/\Gamma_2$, using its anharmonicity $2\chi$ and decoherence rate $\Gamma_2=\kappa/2$.
In this work, the SQUID array resonator is only moderately nonlinear, with $\eta_{\rm SQ}=4|\chisq| / \kappa_\mathrm{SQ} \approx \SI{0.27}{}$.
As a result, the hybridised modes do not reach the quantum regime ($\eta > 1$) and $\eta_i = 4|\tilde \chi_i| / \tilde \kappa_i$ is at most $\SI{0.12}{}$ (\autoref{fig:appendix_qubit_weak_chi}).
Nevertheless, as noted in the main text, this moderate nonlinearity regime is attractive for applications such as parametric processes, frequency conversion and squeezing, and quantum sensing with multimode massive objects.

It was shown that coupling a low loss microwave cavity to a transmon qubit can lead to a cavity with inherited self-Kerr much larger than its dissipation \cite{kirchmair2013observation}.
This principle was recently extended to a bulk acoustic wave resonator \cite{yang2024mechanical} to realise a single mechanical qubit.
We propose to use the architecture introduced in our work to extend this breakthrough to the multimode regime to realise a network of mechanical qubits.

\textbf{Failure of participation ratio analysis.}
As demonstrated in this work, the participation ratio analysis provides quantitative agreement with measurements in the weak nonlinearity regime.
However it is known that this method can fail for small detuning even for moderately anharmonic circuits like transmon qubits.
To be more quantitive, let us consider a transmon qubit with angular frequency $\omega_q$ and anharmonicity $E_\mathrm C/h = -2\chi_\mathrm{tr}/2\pi=\SI{200}{\MHz}$, coupled to a single bosonic mode, with angular frequency $\omega_q$.
The self- and cross-Kerr of the two hybridised modes strongly deviate from participation ratio predictions (\autoref{eq:particip_KerrA}) for detuning $0<\Delta<E_\mathrm C/\hbar$, with $\Delta = \omega_q - \omega_r$.
This regime is known as the straddling regime \cite{koch2007charge, blais2021circuit}, in which higher levels of the nonlinear resonator cause the self- and cross-Kerr to change sign and show more complicated behavior.
In the limit of very large anharmonicity, $|\chi| \gg |\Delta|, g$, the nonlinear resonator can be approximated as a two-level system.
We then retrieve the celebrated Jaynes Cummings model~\cite{blais2021circuit}.
The nonlinearity of the two hybrid modes is then given by~\cite{yang2024mechanical}
\begin{equation}
    \tilde \chi = -\frac{1}{4} \Delta \pm \frac{1}{4}\left(2 \sqrt{\Delta^2+4 g^2}-\sqrt{\Delta^2+8 g^2}\right).
\end{equation}
In this regime, the nonlinearity of the hybrid modes at $\Delta=0$ is independent of the transmon anharmonicity and is instead limited by the coupling $g$.
This is very different from the prediction of participation ratio analysis (\autoref{eq:particip_KerrA}), where at resonance the participation ratio is $\SI{50}{\percent}$ and the self-Kerr of hybrid modes is $\tilde \chi = \chi_\mathrm{tr}/4$, which is independent of $g$.
The intermediate regime, where $|\chi| \simeq |\Delta|,g$, is more complicated but can be addressed using full numerical diagonalization.

\begin{figure}
    \includegraphics[width = 1\linewidth ]{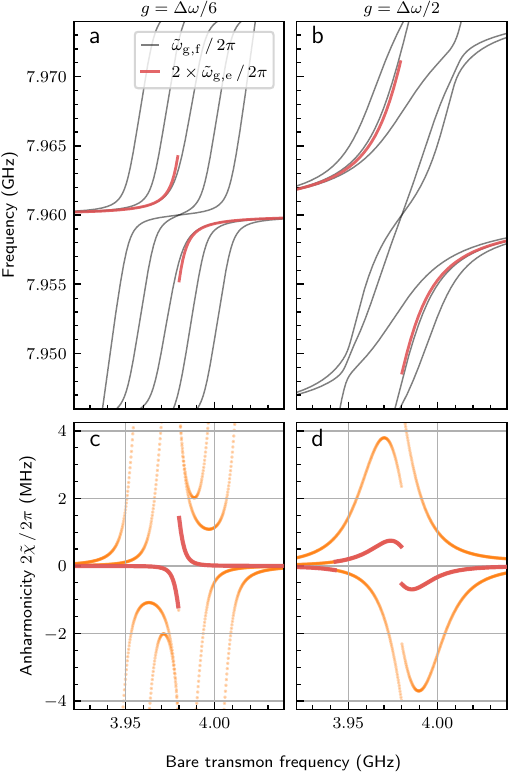}
    \caption{\label{fig:appendix_qubits_2exc_Kerr}
    \textbf{Two excitation subspace and anharmonicities.
    }
    Simulations use the same model as in \autoref{fig:outlook}.
    (a, b)~Part of the two-excitation energy levels $\tilde \omega_{g,f}\,/\,2\pi$ (grey lines) zoomed-in around twice the single ground state transition energy $2\times\tilde \omega_{g,e}\,/\,2\pi$ of the central acoustic mode (red).
    (c, d) Anharmonicities $2\tilde \chi\,/\,2\pi$ associated to the red hybrid mode of \subautoref{fig:outlook}{a,b} within $\SI{\pm 4}{\GHz}$ are shown with orange circle markers.
    The smallest (in absolute value) anharmonicity is shown with a thick red line and corresponds to the anharmonicity shown in \subautoref{fig:outlook}{c,d}.
    }
\end{figure}

\textbf{Numerical diagonalisation.}
To obtain anharmonicities of hybrid system for any resonator nonlinearity, e.g. in the straddling regime, we can resort to exact numerical diagonalisation.
We include three levels for all acoustic modes $\hat a_i$ and transmon qubit $\hat b$, and use qutip~\cite{johansson2012qutip} to find the system eigenvalues and eigenvectors.
In \autoref{fig:outlook} we consider a simplified system of five acoustic modes, equally spaced in frequency with $\Delta \omega /2\pi = \SI{15}{\MHz}$.
The anharmonicity of the transmon qubit is $2\chi_\mathrm{tr}/2\pi = -E_\mathrm C/h = \SI{-200}{\MHz}$, corresponding to a typical transmon qubit~\cite{blais2021circuit}. 
We consider a uniform coupling between transmon and acoustic modes, with a single-mode coupling regime $g=\Delta \omega/6=2\pi\times\SI{2.5}{\MHz}$, and a multimode coupling regime $g=\Delta \omega/2=2\pi\times\SI{7.5}{\MHz}$.
The single-excitation energy levels $\omega_{g,e}$ are reported in \subautoref{fig:outlook}{a,b}.
The two-excitation energy spectrum contains 21 levels, part of which is shown with grey solid lines in \subautoref{fig:appendix_qubits_2exc_Kerr}{a,b}.
The red line shows twice the single excitation energy of the red hybrid mode of \subautoref{fig:outlook}{a,b}, which corresponds mainly to the central acoustic mode with bare frequency \SI{3.98}{\GHz}.
To define the anharmonicity of this hybrid mode, we compute the frequency differences $2\tilde \chi = \tilde \omega_{e,f} - \tilde \omega_{g,e}$ for all two excitation levels $\tilde \omega_{e,f}$, i.e. half the difference between the red line and all grey lines.
We thus get several anharmonicities $2\tilde \chi$ (21 for 5 acoustic modes). The one within $\SI{\pm 4}{\MHz}$ are reported in \subautoref{fig:appendix_qubits_2exc_Kerr}{c,d} with orange markers.
The smallest absolute anharmonicity per bare transmon frequency is shown with a thicker red line, and corresponds to the value shown in \subautoref{fig:outlook}{c,d}.
Some of the transitions may have zero matrix elements, but we consider the most stringent case by including all transitions.

\begin{figure}
    \includegraphics[width = 0.98\linewidth ]{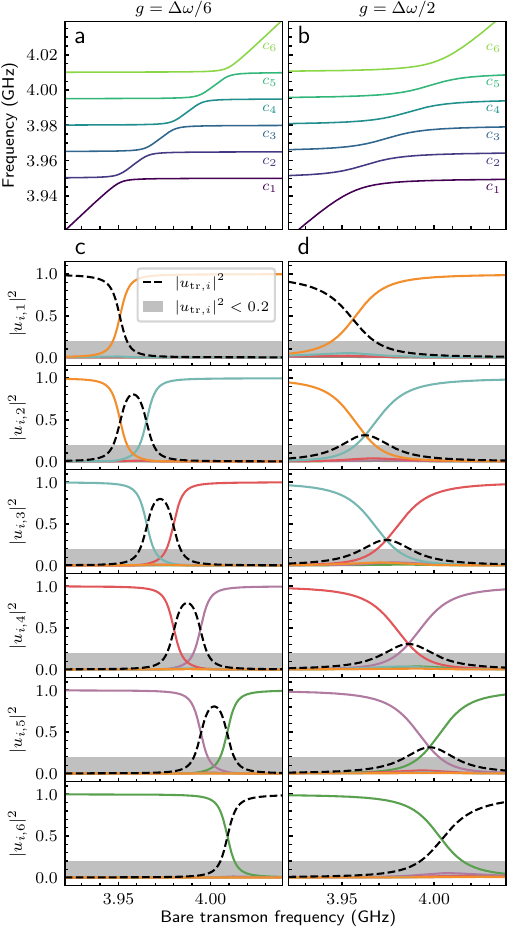}
    \caption{\label{fig:appendix_qubits_all_particip}
    \textbf{Bare modes participation in hybrid qubits.
    }
    Simulations use the same model as in \autoref{fig:outlook}.
    (a,b) The single excitation energy spectrum is reproduced from \subautoref{fig:outlook}{a,b}.
    Each color corresponds to a different hybrid mode from $c_1$ to $c_6$.
    (c,d) hybrid modes decomposition in the bare mode basis.
    Each row corresponds to the decomposition of one hybrid mode of (a,b).
    The bare transmon participation ratio is shown in black dashed lines, while the participation of other bare acoustic modes is shown with coloured solid lines.
    The area shaded in grey corresponds to hybrid modes with transmon participation under $\SI{20}{\percent}$, which we classify as mainly mechanical.
    }
\end{figure}

A hybrid mode can be considered a qubit if its smallest anharmonicity (by magnitude) is much larger than its decoherence rate.
However, to be considered a \emph{mechanical} qubit, it should have a weak bare qubit participation ratio $|u_\mathrm{tr}|^2$.
To identify mechanical qubits in \autoref{fig:outlook}, we decompose the hybrid modes in the bare mode basis, as shown in \autoref{fig:appendix_qubits_all_particip}.
The bare qubit participation is shown with black dashed lines, and we only consider hybrid modes with participation weaker than $|u_\mathrm{tr}|^2 \, < \, 20\%$.

\textbf{Effect of transmon qubit dissipation rate.}
In \autoref{fig:outlook}, we assumed a moderate total dissipation rate for the transmon qubit, $\kappa_\mathrm{tr}/2\pi = \SI{300}{\kHz}$, which is consistent with values reported in the literature for direct fabrication of transmons on GaAs~\cite{sletten2019resolving}. 
However, this value is lower than what we measured for the SQUID array resonator in this work, $\kappa_\mathrm{SQ,int}^\mathrm{extra}/2\pi = \SI{2}{\MHz}$ (\autoref{fig:SQ_dissip_appendix}).
We numerically investigated the impact of the transmon qubit dissipation rate on the number of mechanical qubits, fixing all the others parameters as in \autoref{fig:outlook}.
For $\kappa_\mathrm{tr}/2\pi = \SI{600}{\kHz}$, we still obtain four mechanical qubits over an extended range of bare transmon frequencies. 
For $\kappa_\mathrm{tr}/2\pi = \SI{1000}{\kHz}$, three mechanical qubits remain accessible. 
At $\kappa_\mathrm{tr}/2\pi = \SI{1500}{\kHz}$, only a single mechanical qubit can be realised; however, it is possible to recover four simultaneous mechanical qubits by increasing both the coupling strength $g$ and the free spectral range (FSR).

\begin{figure}
    \includegraphics[width = 1\linewidth ]{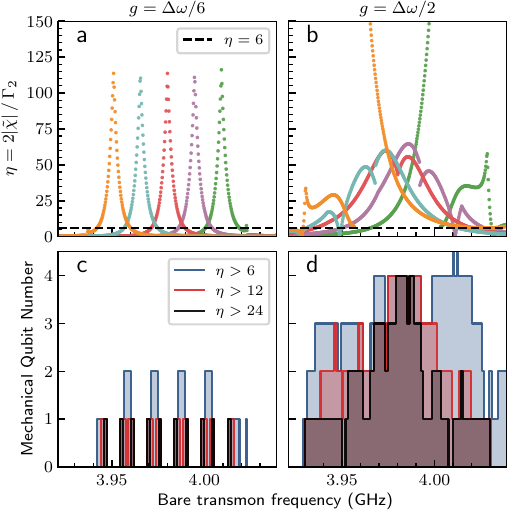}
    \caption{\label{fig:appendix_qubits_flipchip}
    \textbf{Simulations for mechanical qubits with reduced dissipation rates. 
    }
    Simulations use the same model as in \subautoref{fig:outlook}{e-h}, but with reduced total dissipation rates $\kappa_\mathrm{SQ}/2\pi = \kappa/2\pi = \SI{25}{\kHz}$, compatible with a flip-chip architecture.
    Compared to \autoref{fig:outlook}, the nonlinearity to decoherence ratios are larger (a,b), enabling mechanical qubits over a broader range of bare transmon frequencies and with larger nonlinearity to decoherence ratios (c,d).
    }
\end{figure}

The dissipation rate of the transmon qubit can be further reduced by using a flip-chip integration technique~\cite{satzinger2019simple, 
conner2021superconducting, kitzman2023phononic}, where all components except the transmon qubit are fabricated on GaAs, and the qubit is fabricated on a silicon or sapphire substrate.
Such architectures have been experimentally demonstrated, yielding transmon qubit lifetimes of $\SI{\approx10}{\us}$~\cite{conner2021superconducting}.
In \autoref{fig:appendix_qubits_flipchip}, we assume a compatible transmon decoherence rate of $\kappa_\mathrm{SQ} /2\pi = 2 \Gamma_2 /2\pi = \SI{25}{\kHz}$.
We also assume reduced SAW dissipation rates, setting $\kappa/2\pi=\SI{25}{\kHz}$, which has been demonstrated on similar GaAs substrates with optimised IDT and Bragg mirror designs~\cite{andersson2022squeezing}.

\textbf{Optimizing the number of mechanical qubits.}
We note that increasing the coupling $g$ excessively can be detrimental to the realisation of multiple mechanical qubits. 
In fact, in this regime, the transmon participation in the hybrid modes becomes vanishingly small, except for the first and last hybrid modes of the passband, which retain substantial transmon participation
This behaviour is akin to the formation of atom-photon bound states in systems with dispersion relations featuring bandgaps~\cite{calajo2016atom}.
Therefore, the coupling $g$ should be kept smaller than the SAW cavity bandwidth to realise multiple mechanical qubits.

The transmon qubit anharmonicity can also be optimised to maximise the number of mechanical qubits. 
In the regime where $\chi_\mathrm{tr} \gg g$, the anharmonicity of the hybrid modes is primarily determined by $g$. 
However, by carefully selecting $\chi_\mathrm{tr}$ to be moderate, one can access a straddling regime~\cite{koch2007charge} in which the hybrid modes retain significant anharmonicity over a wider range of bare transmon frequencies, potentially increasing the number of mechanical qubits. 
Conversely, if $\chi_\mathrm{tr}$ is too large, the transmon’s second transition frequency $\omega_{e,f}$ is very detuned from the bare SAW mode frequencies, thereby limiting the effectiveness of the straddling regime. 
For the parameters used in \autoref{fig:outlook}, we find that $\chi_\mathrm{tr}/2\pi = -\SI{100}{\MHz}$ offers a good compromise.

\section{Experimental Details: Design, Fabrication, Setup}

\subsection{Acoustic relaxation rate}

Outside the Bragg mirror stopband, the acoustic relaxation rate of the SQUID array resonator is given by~\cite{aref2016quantum, moores2018cavity}
\begin{equation}
    \Gamma(X)= \frac{1.3}{2\sqrt{2}} K^2 N_p \, \omega_0^c\left(\frac{\sin X}{X}\right)^2,
    \label{eq:saw_emission_rate}
\end{equation}
where $K^2=0.07$\,\% is the effective electromechanical coupling in GaAs, $N_p=14$ is the number of split-finger pairs in the coupler IDT, and $X=\pi N_p (\omega-\omega_0^c)/\omega_0^c$, with $\omega_0^c/2\pi$ the central frequency of the SAW cavity and $\omega/2\pi$ the frequency of the electromagnetic mode.
In our case, the coupling results in approximately $\Gamma/N_p \approx 2\pi\cdot \SI{1.3}{\MHz}$ per split-finger pair.

\subsection{Fabrication}
The fabrication begins with a cleansing of an intrinsic semi-insulating gallium arsenide wafer (AXT Inc) in isopropanol (IPA). 

\textbf{Intergidital transducer (IDT)} - The 2" GaAs substrate is spin-coated with a double resist layer of 60\,nm of MMA followed by 100\, nm PMMA. Both layers are baked at 180C for 5 minutes. The crystallographic direction [100] is aligned in the microscope with a precision of $\pm 0.5$ degrees, and mounted on the cassette of an E-beam lithography machine (Raith EBPG 5200). The IDTs are exposed with a beam of 2\,nA and a base dose of 250\,$\mu \rm{C}/\rm{cm}^2$, proximity error correction and with 20\,nm negative bias on the fingers' width. After developing the resist in a solution of MIBK:IPA 1:1 for 60\,s the wafer is blow dry with a nitrogen gun and exposed to 10\,s oxygen plasma generated with 50\,W of power (Plasma-Therm). The sample is loaded in the evaporation chamber of a Plassys evaporator and pumped overnight to reach a base pressure of $5\cdot 10^{-8}$\,mBar. Following the evaporation of 30\,nm thick layer of Aluminium, the wafer is placed in a pre-heated beaker with Rem1165 at 85\,C for 30 minutes and left in lift-off for additional 3 hours. After the aluminium residues have been removed with a pipette, the wafer is transferred to a new beaker with Rem 1165 at 85 C for 5 minutes, followed by sonication at 40\% power at 35\,kHz for 3 minutes. The same operation is repeated after transferring the wafer to room temperature Methanol and finally to a beaker with room temperature IPA. 

\textbf{Ground Plane} - The wafer is spin-coated with double-layer optical resist,  300\,nm of LOR 3A, baked at 180C for 5 minutes and 500\,nm of S1805, baked  at 115C for 5 minutes. The resist is exposed with $97\,$mJ/cm$^2$, developed in MF319 for 60\,s with a light manual agitation and rinsed with a deionised water gun for 2 minutes. After a light descum in Oxygen plasma at 20\,W for 20\,s the wafer is loaded in Plassys evaporator where 150\,nm of aluminium are deposited. For lift-off we follow the same procedure used for the IDT.

\textbf{Josephson junctions} - The wafer is spin-coated with double-resist layer comprising 400\,nm of MMA and 400\,nm of PMMA and both layers are backed at 180\,C for 5 minutes. The exposure of Mahnattan junctions pattern is performed in the EBL with a dose of 600\,$\mu \rm{C}/\rm{cm}^2$, and is followed by a development of 60\,s in  a solution of MIBK:IPA 1:1 and a light descum of in oxygen plasma fat 50\,W for 20\,s. 
The wafer is loaded in a Plassys evaporator where 50\,nm of aluminium is deposited at 45 degrees angle, followed by a static oxidation at 2\,mBar for 20 minutes. 
After rotating the wafer axially by 90 degrees, another deposition of 110\,nm of aluminium is performed. 
Before venting the loadlock, the wafer is left 10 minutes in 10\,mBar oxygen atmosphere. 

\textbf{Patches} - To patch the junctions and the IDT to the ground plane we coat, expose and develop the pattern with the same procedure followed for the junctions. In the evaporator the wafer is exposed to Argon milling at 400\,V for 2 minutes, followed by an evaporation of 200\,nm of aluminium and by an oxidation in 10\,mBar atmosphere for 10\,min. 
Finally the standard lift-off procedure is performed.

As a final remark, the fabrication of the measured sample was performed in October 2020. From measurements performed on nominally identical samples right after the fabrication, we noticed negligible aging in the chip performance. However, holes of hundreds of $\mu\rm m^2$, clearly visible by optical microscopy, appeared on the large aluminum ground-pad.

\subsection{Measurement Setup}
\label{sec:appendix_meas_setup}
\begin{figure}
    \includegraphics[width = 1\linewidth ]{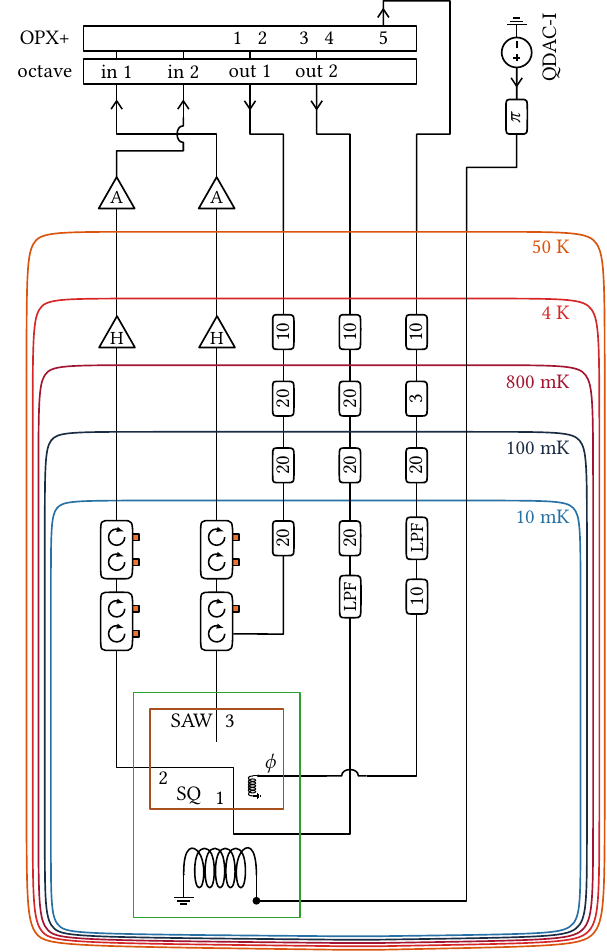}
    \caption{\label{fig:wiring}
    \textbf{    Room-temperature measurement setup and cryostat wiring and filtering.
    }
    Representative measurement setup used in cooldown 20240913CD.  
    Minor changes performed in the other cooldowns are described in \autoref{sec:Attenuation_estimation}.
    }
\end{figure}

The measurement setup is shown in \autoref{fig:wiring}. 
The sample is wire bonded with aluminum wires on a custom printed circuit board and glued on a copper sample holder using PMMA resist.
The sample holder is thermally anchored to the mixing chamber plate of a LD Bluefors cryostat with base temperature \SI{\approx 10}{\milli\kelvin}.
The sample holder is isolated from its electromagnetic environment via a shield with three layers: copper, aluminum and cryoperm.

Measurement signals are generated using a Quantum Machines platform comprising an arbitrary waveform generator (OPX+) and a frequency converter (octave).
The output lines 1 and 2 of the octave are connected to the device via respectively the port 3 (launcher IDT feedline) and port 1 (SQUID array resonator feedline input) with nominally \SI{70}{\dB} attenuation.
The transmission through the SQUID array feedline is collected from port 2 of the device and is amplified by a HEMT LNF-LNC4 8C at the \SI{4}{\kelvin} stage.
The reflected SAW feedline signal is measured using a circulator LNF-CICIC4 8A, and is also amplified by a HEMT LNF-LNC4 8C at the \SI{4}{\kelvin} stage.
A Marki FLP-0490 4.9 GHz low-pass filter is used on the port 1 of the device.
Two double-isolators LNF-CICIC4 8A are used between device and HEMT.
The signal is further amplified at room temperature using an Agile AMT-A0284 amplifier, and demodulated and downconverted in the Quantum Machines system.

The flux line coupled to the SQUID array is attenuated by \SI{43}{\dB}.
A low-pass filter minicircuits  VLFX-225+ is used to further isolate the device from noise near its resonance frequency of $\approx$ \SI{4}{\giga\hertz}.
\SI{10}{\dB} of attenuation is used between the device and the low-pass filter to damp standing wave modes.
The flux line is not used in this work.

Throughout this work, the flux in the SQUID array is controlled using a superconducting coil attached on the bottom of the sample holder.
We use a QDevil QDAC-I voltage source with a $\SI{1}{\kohm}$ low-pass $\pi$-filter with \SI{30}{\hertz} cutoff frequency to apply direct current in the coil.

\subsection{Attenuation estimation}
\label{sec:Attenuation_estimation}

\begin{figure}
    \includegraphics[width = 1\linewidth ]{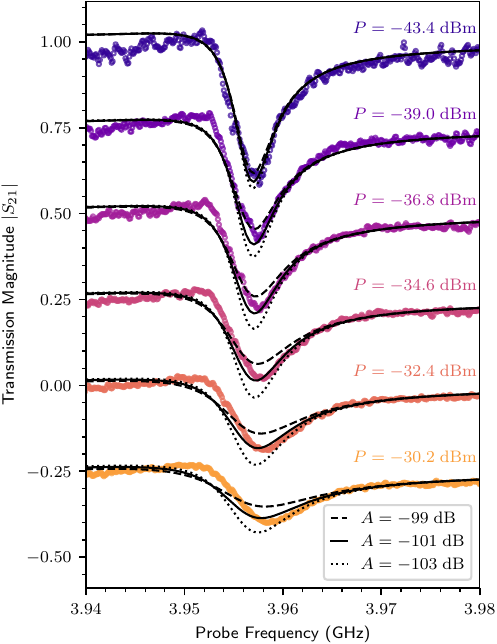}
    \caption{\label{fig:attenuation_calib}
    \textbf{
    Input line attenuation calibration.
    }
    We perform a power sweep on a different device to calibrate the attenuation in the input line of the SQUID array resonator (port 1).
    The device is a Kerr resonator with nonlinearity exceeding its total dissipation rate studied in another work (device N=10 in \cite{peyruchat2025landau}).
    The coloured markers indicate the magnitude of $S_{21}$ measured for different powers $P$ indicated at instrument level outside of the cryostat.
    Each $S_{21}$ trace is shifted along y-axis by $-0.25$.
    We use the method described in \cite{peyruchat2025landau} and fit the power sweep to a master equation model using QuTiP~\cite{johansson2012qutip}.
    We find an attenuation of $-101 \pm 1$ dB (black solid line).
    Dashed and dotted black lines indicate simulation results with $\pm 2$ dB attenuation difference and show a worse agreement with the measurements.
    }
\end{figure}

The measurement data reported in this work were measured across several cooldowns (CD). 
In one cooldown (CD 20241206), we performed a calibration of the input attenuation line of the SQUID array resonator (port 1) using another device connected on a shared cryogenic switch at the mixing chamber stage of the cryostat.
It is a Kerr resonator in the quantum regime, i.e. with nonlinearity larger than dissipation.
The nonlinearity is known from the measurement of multiphoton transitions at high probe power~\cite{peyruchat2025landau}.
We fit the power sweep shown in \autoref{fig:attenuation_calib} with the input attenuation as the only free parameter.
The Kerr resonator device and input power calibration method are explained in greater details in \cite{peyruchat2025landau}.
This allows us to find the self-Kerr $\chisq$ of the SQUID array resonator presented in this work as reported in \autoref{fig:CK}.
The input line connected to port 3 (launcher IDT) is nominally identical to the one connected to port 1.
We determine the attenuation of the launcher IDT input line by using the attenuation calibrated for port 1, accounting for difference in filtering and cabling at the mixing chamber stage of the cryostat using nominal room-temperature values for the different microwave components.

We recalibrate the attenuation of the input lines for all the other cooldowns, according to the following procedure:

\textbf{CD 20241206;} During this cooldown, port 1 (input of the SQUID array resonator feedline) is connected on the same switch as the Kerr resonator \cite{peyruchat2025landau}.
The input attenuation is calibrated using the method described above, and we find $\rm A_3^{\rm SQ}=\SI[separate-uncertainty = true]{-101(1)}{\dB}$.

In the other cooldowns, no switch was used on the SQUID array resonator input (port 1), and the estimate of the attenuation of the input lines is based on the calibration performed during CD 20241206, given that the lines are nominally the same.

\textbf{CD 20240913;} We estimate the attenuation in the launcher IDT feedline by subtracting the nominal room temperature attenuation of the additional components used in CD 20241206, namely
\begin{itemize}
    \item one Marki LPF \SI{4.9}{\giga\hertz} filter (\SI{0.7}{\dB}),
    \vspace{-5pt}
    \item one RLC Electronics F-30-8000 (\SI{0.7}{\dB}), 
    \vspace{-5pt}
    \item one Radiall r591723605 switch (\SI{0.3}{\dB}),
    \vspace{-5pt}
    \item extra cabling (\SI{0.5}{\dB}).
\end{itemize}
Moreover, in CD 20241206 an extra \SI{20}{\dB} internal attenuation is applied in the measurement instrument.
On the whole, we estimate a difference in attenuation of \SI{22.2}{\dB}, which we add to $\rm A_3^{\rm SQ}$ to obtain $\rm A_1^{\rm IDT}=\SI[separate-uncertainty = true]{-78.8(20)}{\dB}$, where the increased error accounts for the misestimation of the attenuation of the different components.

Similarly, we estimate the attenuation of the different components on the SQUID array resonator feedline, finding $\rm A_1^{\rm SQ}=\SI[separate-uncertainty = true]{-78.8(20)}{\dB}$.

\textbf{CD 20240927;} 
Here, to calibrate the input attenuation of the SQUID array resonator line, we perform a power sweep of the SQUID array resonator near zero-flux.
Then, we simultaneously fit several $S_{21}$ traces for increasing power, assuming the value of self-Kerr of the bare SQUID array resonator obtained from the cross-Kerr study (\autoref{fig:CK}).
We find an attenuation $\rm A_2^{\rm SQ}=\SI[separate-uncertainty = true]{-97(1.5)}{\dB}$, where the increased error accounts for the non-idealities in the calibration process.

\begin{figure*}[]
    \includegraphics[width = 1\linewidth ]{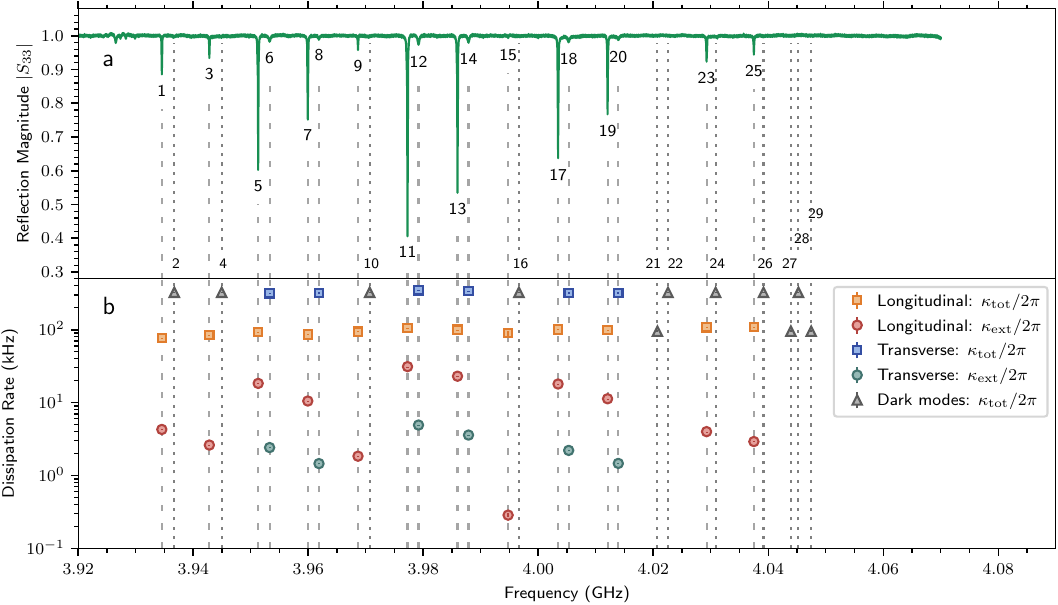}
    \caption{\label{fig:saw_dissip_appendix}
    \textbf{
    Frequency and dissipation rates of bare SAW modes.
    }
    (a) Same trace $S_{33}$ shown in \autoref{fig:bare_saw}.
    The SQUID array resonator is set to zero-flux point with frequency 4.7 GHz.
    (b) In \autoref{fig:bare_saw} we only reported the parameters of longitudinal SAW modes measured from the launcher IDT side ($S_{33}$).
    Here we also report the parameters of transverse SAW modes (blue square and blue-green circle markers) with larger internal dissipation rates.
    Other SAW modes could be identified only from the SQUID array resonator side ($S_{21}$).
    We set their dissipation rates to the average value of the other modes that we could resolve directly (grey triangle markers).
    We use these parameters of the bare SAW modes throughout the rest of the articles.
    The power of the probe tone at device level was \SI{-131}{\dB}m.
    }
\label{fig:bare_SAW_dissip_appendix}
\end{figure*}

Finally, we obtain the attenuation of the feedline of the IDT by estimating the attenuation of the different components compared to the SQUID array resonator line.
The launcher IDT input line has one extra circulator (LNF-CICIC4 8A, \SI{0.2}{\dB}) and  some extra cabling (\SI{0.5}{\dB}).
The SQUID array feedline has an additional low-pass filter (MARKI FLP-0490, \SI{0.7}{\dB}).
Thus, we estimate the attenuation on the IDT feedline to be $\rm A_2^{\rm IDT}=\SI[separate-uncertainty = true]{-97(2)}{\dB}$, where the increased error accounts for the misestimation of the attenuation of the different components.

\textbf{CD 20250214;} we apply the same protocol as in CD 20240927.
We find $\rm A_4^{\rm SQ}=\SI[separate-uncertainty = true]{-102(1.5)}{\dB}$ for the feedline of the SQUID and $\rm A_4^{\rm IDT}=\SI[separate-uncertainty = true]{-103.6(20)}{\dB}$ for the feedline of the IDT.

The attenuation values estimated in the the different cooldowns are summarised in \autoref{tab:Device_setup_parameters}.

\section{Measurement Details}

\subsection{Bare Surface Acoustic Waves}

\subsubsection*{Dissipation analysis}

The SAW modes come in pairs, with one longitudinal and one transverse mode.
Longitudinal modes are strongly coupled to the launcher IDT, which translates into larger $\kappa_\mathrm{ext}$.
The transverse modes have smaller $\kappa_\mathrm{ext}$ and larger $\kappa_\mathrm{int}$.
For convenience, we indexed all modes so that odd numbers are for longitudinal modes and even numbers are for transverse modes.

We characterise the bare SAW modes by setting the SQUID array resonator frequency far detuned from the SAW modes, for example at zero-flux bias for which $\wsq /2\pi \approx \SI{4.7}{\giga\hertz}$.
We measure the reflection signal $S_{33}$ using the setup reported in \autoref{fig:wiring} to obtain the SAW modes bare frequencies and dissipations.

In \autoref{fig:bare_saw} of the main text, we only report the longitudinal modes that we could resolve from port 3 (launcher IDT).
In \autoref{fig:saw_dissip_appendix} we also report the parameters of the transverse modes extracted from port 3 (blue and blue-green markers).

In addition, we could also identify 11 extra SAW modes by flux tuning the SQUID array resonator across the SAW modes, and measuring ports 1 and 2 ($S_{21}$ measurements, see \autoref{fig:fluxsweep}).
These SAW modes are coupled to the SQUID array resonator but are not coupled to the launcher IDT. 
The difference in coupling is due to the different placement of the launcher and coupler IDTs along the Bragg mirror.
We do not have a direct access to the dissipation rates of these extra dark modes. 
We set their total dissipation rates to the average of the other longitudinal (\SI[separate-uncertainty = true]{95}{\kilo\hertz}) or transverse modes (\SI[separate-uncertainty = true]{326}{\kilo\hertz}), shown as grey markers in \autoref{fig:saw_dissip_appendix}.
These values are used in \autoref{eq:hybrid_kappa} to perform the global fit of $\kappa_\mathrm{SQ}$ in \autoref{fig:participation}.

\subsubsection*{SAW mode spacing (FSR)}
\begin{figure}
    \includegraphics[width = 1\linewidth ]{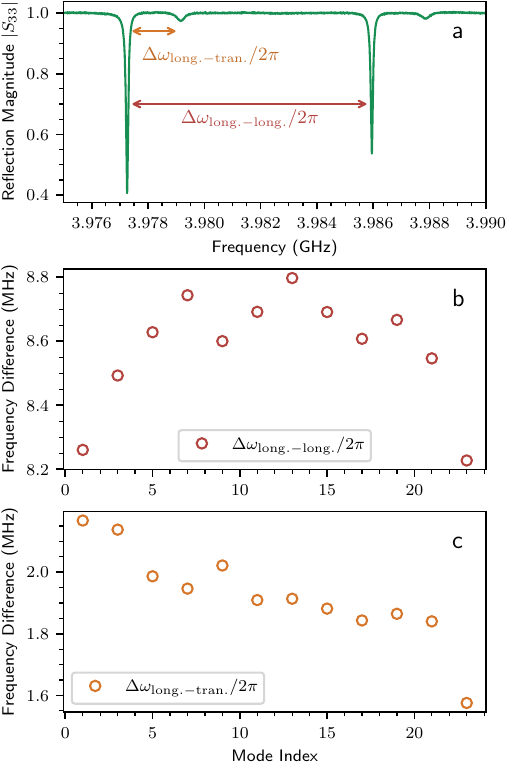}
    \caption{\label{fig:saw_fsr_appendix}
    \textbf{
    Frequency difference between SAW modes.
    }
    (a) Zoom-in on the trace $S_{33}$ of bare SAW modes reported in \autoref{fig:bare_saw}~(a).
    (b,c) From the SAW mode positions shown in \autoref{fig:saw_dissip_appendix}, we compute the frequency difference between neighboring longitudinal SAW modes (b) and between a longitudinal mode and its neighboring transverse mode (c).
    }
\end{figure}

The free spectral range (FSR) of the SAW resonator is approximately equal to \SI{8.6}{\mega\hertz}, and it is determined by the effective cavity length, $L_c$.
The FSR is shown in \autoref{fig:saw_fsr_appendix}.
Longitudinal SAW modes are separated by 8.25 to 8.8 MHz, with the largest deviation on the edge of the mirror bandwidth due to increasing effective cavity length \cite{sletten2019resolving}.
The transverse modes have a resonance frequency from \SI{1.6}{} to \SI{2.15}{MHz} higher than the longitudinal modes~\cite{fisicaro2025imaging}.

\subsubsection*{Dissipation versus power}

\begin{figure}
    \includegraphics[width = .95\linewidth ]{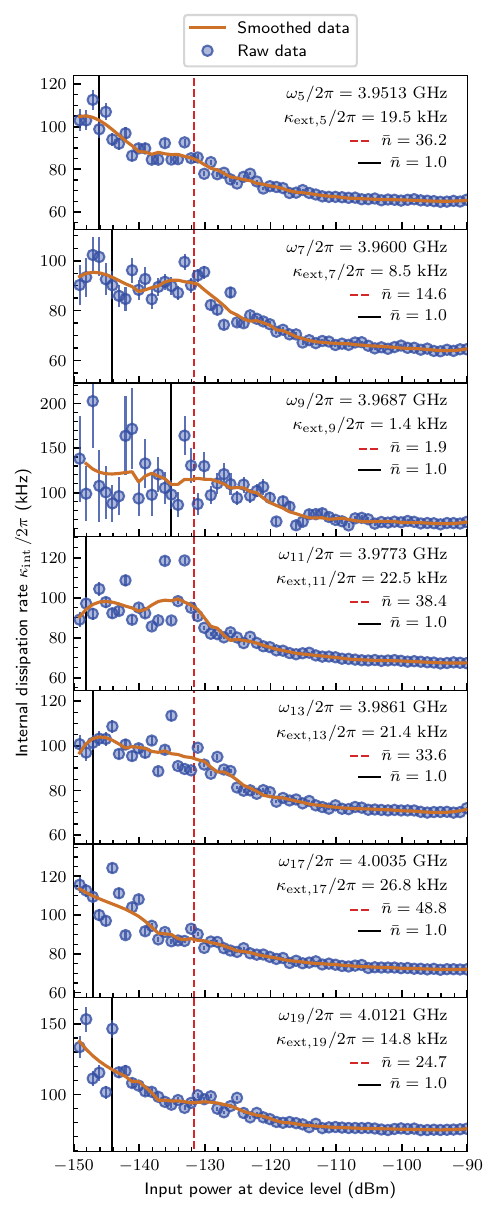}
    \caption{\label{fig:saw_dissip_power_appendix}
    \textbf{Internal dissipation rate of SAW modes as a function of input power.}
    We measure $S_{33}$ for increasing probe power and extract the dissipation rates by fitting each SAW mode.
    Dissipation rates extracted are shown with blue markers, and the trend smoothed with a savgol filter is shown with a solid orange line.
    External dissipation rates do not depend on power, and we report their values in each subplot.
    The input power is converted to average phonon number $\bar n$. 
    The power corresponding to $\bar n = 1$ is shown with solid black lines, and the power used in \autoref{fig:bare_saw} with dashed red lines.
    } 
\end{figure}

We investigate the dependence of the dissipation of the bare SAW modes on the power of the input field.
This measurement was performed with a vector network analyser during the last cooldown reported in \autoref{tab:Device_setup_parameters}.
The attenuation of the line is obtained by fitting a power sweep of the SQUID array, as detailed above \autoref{sec:Attenuation_estimation}.
As shown in \autoref{fig:saw_dissip_power_appendix}, the total dissipation decreases for increasing power, typical of the saturation of spurious two-level systems.
The red dashed line corresponds to the power at which the trace shown in the main text \autoref{fig:bare_saw} and in \autoref{fig:bare_SAW_dissip_appendix} was acquired.

\subsection{SQUID Array Dissipation}
\label{app:SQUID_array_dissipation}
\begin{figure*}
    \includegraphics[]{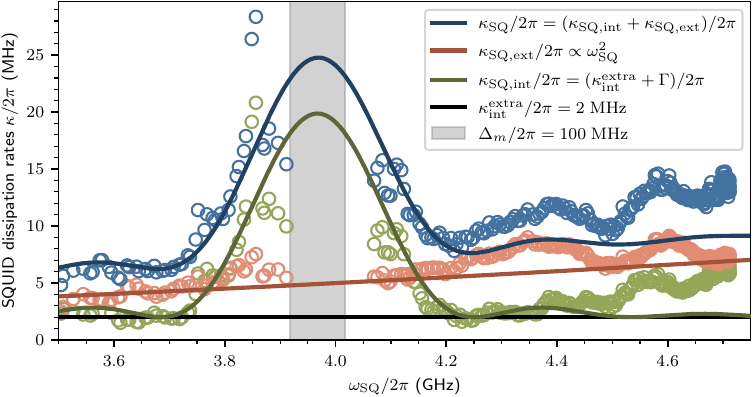}
    \caption{\label{fig:SQ_dissip_appendix}
    \textbf{
    Dissipation of SQUID Array versus Flux
    }
    We sweep the flux threading the SQUID array resonator, measure $S_{21}$ and fit the traces with \autoref{Eq:Notch_coefficient_single_mode_FANO} to extract the dissipation rates.
    Extracted values of internal (orange), external (blue) and total (green) dissipation rates are shown with markers.
    The internal dissipation rate from phonon emission via the coupler IDT ($\Gamma$, \autoref{eq:saw_emission_rate}) is enhanced near the central frequency of the SAW cavity $\omega_0^c/2\pi\approx \SI{3.967}{\giga\hertz}$.
    Additional internal losses are estimated to be approximately constant at $\kappa_\mathrm{int}^\mathrm{extra}/2\pi \approx \SI{2}{\MHz}$ near the mirror bandwidth (black line).
    SQUID array dissipation cannot be directly measured within the mirror bandwidth, (gray shaded region astride $\omega_0^c/2\pi$).
    The power of the probe tone was \SI{-142}{\dB}m, corresponding to an average photon population in the SQUID array of $\bar{n}_{\rm SQ} = \SI{0.08}{}$. 
    }
\end{figure*}

We report here the measured dissipation rates of the SQUID array resonator as a function of the bare SQUID array resonator frequency, changed via magnetic flux generated by an external superconducting coil.
This measurement is performed using a vector network analyzer.
For each flux value, we measure $S_{21}$ and fit it to \autoref{Eq:Notch_coefficient_single_mode_FANO} to obtain internal and external dissipation rates.
The results are shown in \autoref{fig:SQ_dissip_appendix}.

The external dissipation rate of the SQUID array resonator is determined by the capacitive coupling to the $Z_0=50 \; \Omega$ feedline.
The frequency of the SQUID array resonator is tuned by changing the effective Josephson inductance using magnetic flux. 
We thus expect that when changing the resonance frequency of the SQUID array resonator, its external coupling rate scales as 
\begin{equation}
    \kappa_\mathrm{ext,SQ} = \frac{ C_{\rm c}^2}{ C_{\rm tot}}\, Z_0 \ \wsq^2 = \rm \tau \, \wsq^2\ ,
\end{equation}
where $C_{\rm c}$ si the coupling capacitance and $C_{\rm tot}$ the total capacitance of the SQUID array resonator. 
The solid green line in \autoref{fig:SQ_dissip_appendix} follows this formula with a value of $\tau = \SI{0.31}{\pico\second}$ chosen to match the measured external dissipation rate around the mirror bandwidth.

Ripples in dissipation rates are observed for $\wsq > \SI{4.2}{\GHz}$.
External dissipation rate is increased near \SI{4.4}{\giga\hertz} and \SI{4.6}{\giga\hertz}, while internal dissipation rate seems to increase near \SI{4.55}{\giga\hertz} and \SI{4.7}{\giga\hertz}.
The origin of these ripples might be related to the presence of standing waves from spurious modes in the electromagnetic environment.

The internal dissipation rate can originate from two mechanisms.
First, we estimate the spurious internal loss rates to be roughly constant from \SI{3.6}{\giga\hertz} to \SI{4.3}{\giga\hertz} with about $\kappa_\mathrm{int}^\mathrm{extra} / 2\pi \approx \SI{2}{\MHz}$.
Second, the SQUID array can emit phonons in the substrate via the coupler IDT at its capacitive end.
This coupling $\Gamma$ varies with the frequency following a $\rm sinc^2$ profile as detailed in \autoref{eq:saw_emission_rate}.
However, the Bragg mirrors open a gap in the phonon continuum of the substrate. 
Therefore, we expect this second emission channel of the SQUID array to be suppressed inside the mirror stop band (grey shaded region in \autoref{fig:SQ_dissip_appendix}), except at the frequency of the bare cavity modes~\cite{sletten2019resolving}.

\subsection{Hamiltonian Parameter Extraction}
\label{subsec:appendix_multimode_fit}

We perform a global fit of the linear RWA Hamiltonian \eqref{Eq:RWA_hamiltonian_matrix} to the resonant flux sweep shown in \subautoref{fig:fluxsweep}{b, c}. 
Namely, for each flux value we extract the frequency of the resolved eigenmodes for both $S_{21}$ and $S_{33}$ traces.
The loss function is computed globally over the entire 2D map, using the difference between the minima extracted and the eigenvalues predicted by the numerical diagonalisation of the matrix $T$ (\autoref{Eq:T_and_U_matrices}).
The free parameters of the fit are the frequencies of the $11$ SAW modes which cannot be resolved from the spectroscopy (grey markers in \autoref{fig:bare_SAW_dissip_appendix}), together with the 29 coupling strengths between the electromagnetic and mechanical modes, $g_i$.

The fitting protocol is performed iteratively in two steps: a first general fit is performed over the full bandwidth of the mechanical cavity as shown in \autoref{fig:fluxsweep}, and then a second fit is implemented over a narrower region across the bare SAW modes $[\omega_8; \omega_{16}]$.
The second fitting step aims at refining the extraction of the parameters for the participation and the dissipation studies presented in \autoref{fig:participation}.
In the refined fit, the free parameters are the frequency of mode $b_{16}$ (dark form port 3) and all the coupling strengths $[g_{8}; g_{16}]$.
Finally, to keep consistent numbering between bare and hybridised modes, we set the couplings of the bare SAW modes $b_{9}, \ b_{10}, \ b_{24}$ to zero.   

\subsection{Derivative and Dissipation of hybridised modes}
\label{subsec:appendix_deriv_dissip}

\subsubsection*{Additional data derivative and dissipation}

\begin{figure*}
    \includegraphics[width = 1\linewidth ]{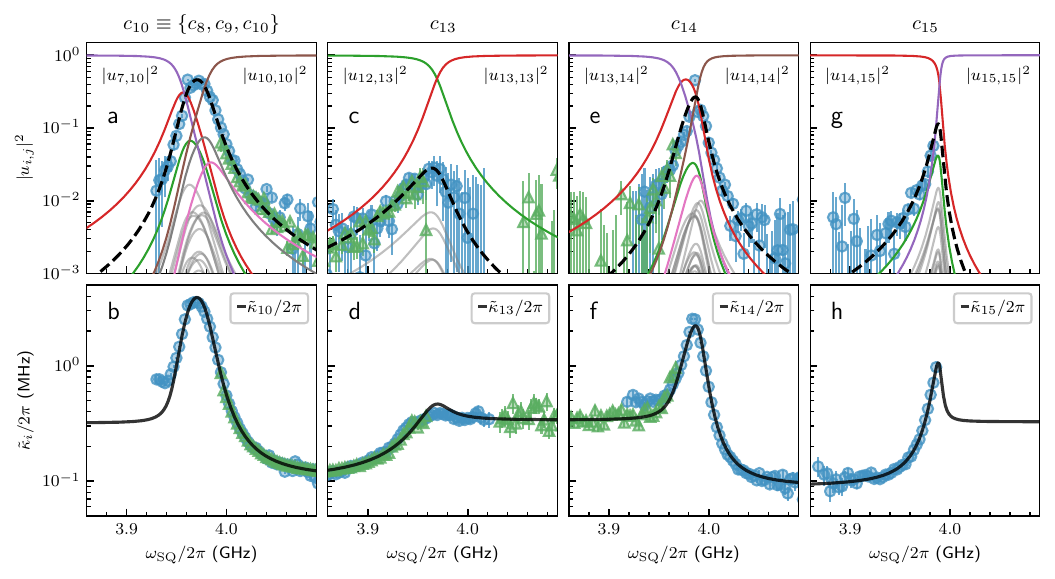}
    \caption{\label{fig:deriv_dissip_appendix}
    \textbf{
    Additional data for \autoref{fig:participation}.
    }
    Bare modes participation ratios (top row) and total dissipation rates (bottom row) for four additional hybridised modes, $\{c_{10}, c_{13}, c_{14}, c_{15}\}$.
    The analysis is performed on the same detailed flux sweep as shown in \subautoref{fig:participation}{a,b}.
    }
\label{fig:part_dissip_extended_data}
\end{figure*}

The protocol to extract the participation of the SQUID array resonator from the derivative of the measured frequency of the hybrid modes is applied to the six hybrid modes in the range from $c_{10}$ to $c_{15}$.
The two bare SAW modes $b_9$ and $b_{10}$ have vanishing coupling to the SQUID array resonator, $g_9=g_{10}=0$.
Therefore, the three hybrid modes $c_8$, $c_9$ and $c_{10}$ appear as a single continuous mode. 
It is the pumped mode $c_p$ shown with pink dotted lines in \subautoref{fig:CK}{a}, and in this section we refer to it as $c_{10}$ for simplicity.
The extended data set of \autoref{fig:participation} is displayed in \subautoref{fig:part_dissip_extended_data}{a, c, e, g}, showing good agreement both from reflection and transmission data, in blue and green respectively. Modes with maximum participation below \SI{2}{\percent} are shown as solid grey lines.
The analysis of the dissipation is reported in \subautoref{fig:part_dissip_extended_data}{b, d, f, h}, where the solid black line is the result of the simultaneous fit to \autoref{eq:hybrid_kappa} over all the modes under study, including the ones shown in \autoref{fig:participation}.

A more comprehensive model of the total dissipation rate of the hybrid modes would account for interference between the external coupling rates of the bare SAW modes to the IDT, as detailed in \cite{sletten2021quantum}.
However, in our device, the total dissipation rate of the bare SQUID array resonator is around two orders of magnitude higher than the one of the bare SAW modes.
Therefore, when modes are hybridizing, their dissipation rate $\tilde \kappa$ is dominated by the bare SQUID array resonator contribution, so we can safely neglect possible interference effects.

The error reported on the value of $\kappa_{\rm SQ}$, extracted from the simultaneous fit, is increased to account for the uncertainty on the participation ratios, $|u_{ij}|^2$, computed from the electromechanical coupling strengths $g_i$ and the frequency of the bare SAW modes $\omega_i$.

\subsubsection*{Extra spurious peak}

\begin{figure*}
    \includegraphics[]{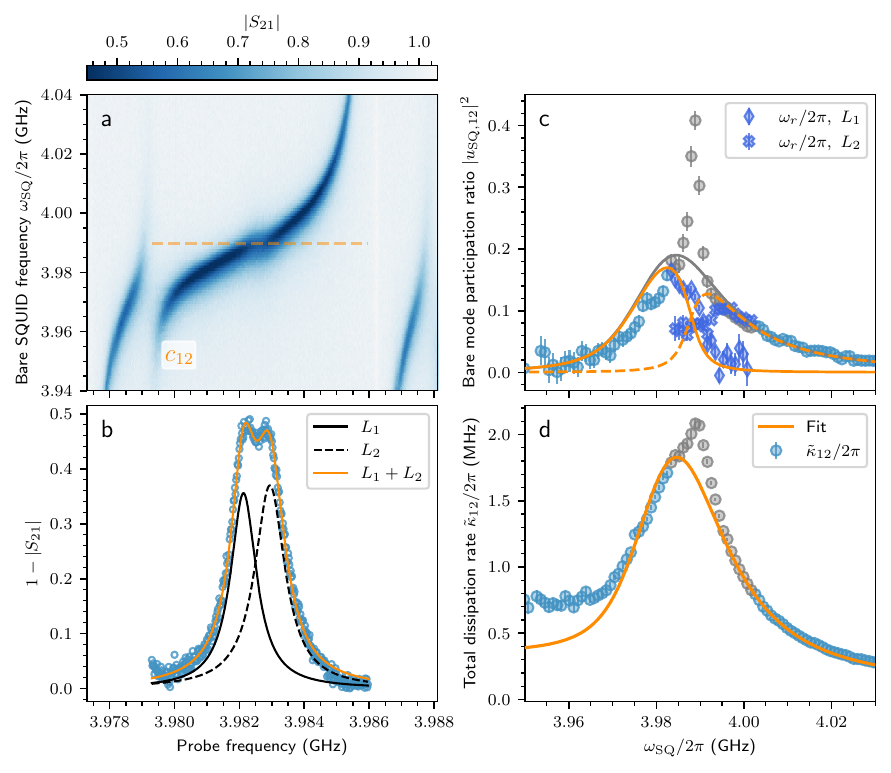}
    \caption{\label{fig:double_peak_appendix}
    \textbf{
    Mode participation and dissipation rate analysis in presence of a spurious SAW mode.
    }
    (a) Zoom-in view of the flux sweep shown in \subautoref{fig:participation}{a}, highlighting the presence of an additional SAW mode of bare frequency $\approx 3.983$ GHz.
    (b) We report $1-|S_{21}|$ at the bare SQUID array resonator frequency $\omega_\mathrm{SQ}/2\pi=3.99$ GHz (orange dashed line in (a)), where the hybrid mode $c_{12}$ is maximally hybridised with the spurious SAW mode.
    Measurement data are shown as blue markers.
    The data are fit to a double Lorentzian model (\autoref{eq:double_lorentzian_model}), yielding the solid orange line.
    The individual Lorentzian curves $L_1$ and $L_2$ are shown as black solid and dashed lines, respectively.
    (c) Participation of the bare SQUID array resonator in the hybridised mode $c_{12}$ and spurious mode.   
    Simulated participation ratios are shown for the case without the spurious mode (grey solid line) and with the spurious mode included (orange solid and dashed lines).  
    The SQUID array resonator participation ratio is also obtained from the derivative of the measured eigenfrequencies (\autoref{eq:particip_ratio_deriv}).
    Circle markers come from the same protocol described in \autoref{fig:participation}, which fails in the presence of a spurious SAW mode (grey markers).
    Cross and diamond blue markers are obtained from the derivative of the position of Lorentzians $L_1$ and $L_2$.
    (d) Dissipation rate of hybridised mode $c_{12}$.
    Circle markers are obtained from measured $S_{21}$ using the protocol of \autoref{fig:participation}.
    The dissipation is over-estimated in the double peak region (grey) compared to the solid orange line reporting the result of the global fit of the total dissipation rates of the hybrid modes (\subautoref{fig:participation}{d, f}).
    }
\end{figure*}

We observed several weakly coupled spurious modes. 
To accurately extract the participation ratio of the SQUID array resonator based on the derivatives of the hybrid modes, we first characterised how the SQUID array resonator couples to these spurious modes. 
These modes do not follow the pattern of longitudinal and transverse modes reported in \autoref{fig:bare_SAW_dissip_appendix}.
These spurious modes are possibly associated with higher-order transverse excitations and appear as small avoided crossings, as shown in \subautoref{fig:double_peak_appendix}{a,b} for the hybrid mode $c_{12}$.
To track the position of these features, we fit the spectrum at each bare SQUID array resonator frequency using a double Lorentzian model:
\begin{equation}
    L_1(\omega) + L_2(\omega)= \frac{A_1}{1 + \left( \frac{\omega - \omega_{r,1}}{\sigma_1} \right)^2} + \frac{A_2}{1 + \left( \frac{\omega - \omega_{r,2}}{\sigma_2} \right)^2} \ , 
\label{eq:double_lorentzian_model}
\end{equation}
where $A_i$ are normalisation factors, $\omega_{r, i}$ are the resonance frequencies, and $\sigma_i$ are the half-widths at half-maximum for each Lorentzian.
The coupling between the SQUID array resonator and the spurious mode is estimated by applying the same fitting protocol detailed in \autoref{subsec:appendix_multimode_fit} to the frequency of the two resonances extracted from the double Lorentzian fit, which yields $g_{spurious} /2\pi \approx \SI{0.8}{\mega\hertz}$. 
Then, we take the derivative of the position of the resonance frequency of each Lorentzian with respect to the bare SQUID array resonator frequency (blue diamonds and crosses in \subautoref{fig:double_peak_appendix}{c}).
We find good agreement with the participation of the SQUID array resonator extracted from the numerical diagonalisation of the linear model complemented with the extra peaks (orange lines).

In practice, we always use the single-peak model to extract the dissipation of the hybrid modes. 
However, when the data can be successfully fit with a two-peak model (see \autoref{eq:double_lorentzian_model}), the single-peak model fails to provide an accurate dissipation value. 
In these cases, we exclude the corresponding data from our dissipation analysis and use the results of the two-peak fit solely to define the exclusion regions. 
This procedure prevents overestimation of dissipation, as indicated by the grey markers in \subautoref{fig:double_peak_appendix}{d}.

\subsection{Self-Kerr Measurements}
\begin{figure*}
    \includegraphics[width = 1\linewidth ]{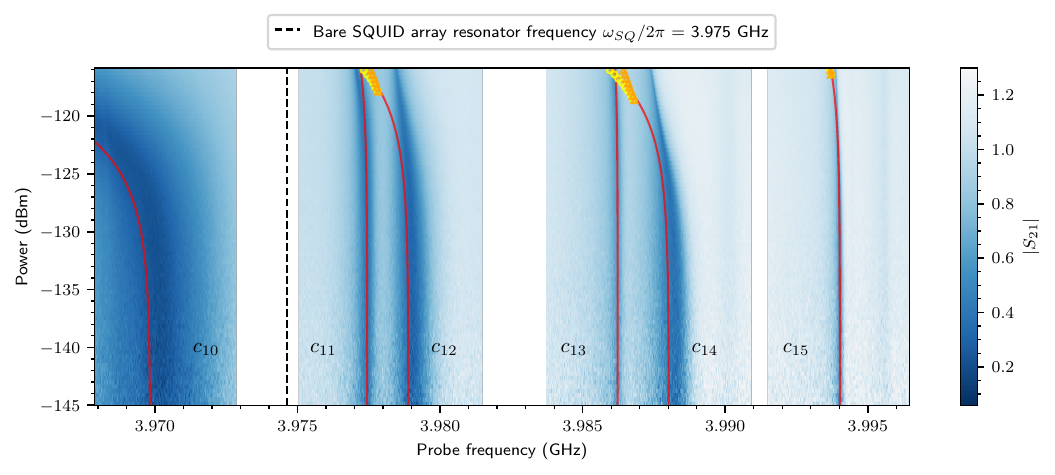}
    \caption{\label{fig:self_kerr_appendix}
    \textbf{
    Self-Kerr measurement at one flux value.
    }
    Magnitude of the normalised transmission signal $|S_{21}|$ for increasing power of the probe tone. 
    The measurement was performed around each hybrid modes individually.
    Red solid line shows the results of simulations for a driven classical Duffing resonator assuming the modes behave as independent oscillators.
    No extra fit parameters are used.
    The dashed black line marks the frequency of the bare SQUID array resonator $\wsq$.
    Yellow and orange markers show the region where each oscillator undergoes a bifurcation and enters a metastable region.
    }
\end{figure*}

The protocol to measure the self-Kerr of the hybrid modes consists in  a resonant single-tone power sweep across the modes of interest.
In \autoref{fig:self_kerr_appendix}, the self-Kerr map of the hybrid modes $c_{10}$ to $c_{15}$ 
is compared to the simulation of a driven classical Duffing oscillator as a function of the amplitude of the drive~\cite{eichler2014controlling, peyruchat2025landau}.
In the simulation all the parameters are fixed, and in particular the self-Kerr inherited by the SAW modes is derived from \autoref{eq:particip_Kerr}.
The theoretical curves of the model qualitatively capture the displacement of the SAW modes, until either two SAW modes start to repel each other or the mode enter the metastable region shown by the yellow and orange markers.

\subsection{Cross-Kerr Measurements}
\label{subsec:CK_meas_protocol}

\subsubsection*{Measurement protocol}

Cross-Kerr coefficients, $\tilde{\chi}_{i,j}$, are a quadratic correction to the linear model, which quantifies the shift in the frequency of hybrid mode $c_{i}$ as a function of the average number of excitations in hybrid mode $c_{j}$.
The protocol that we adopted for measuring cross-Kerr coefficients consists of two-tone spectroscopy measurements, where we apply a pump tone on mode $c_j$ and a weak probe tone on $c_j$ \cite{muppalla2018bistability}.
The need to sweep the pump tone arises from the self-Kerr nonlinearity acquired by the hybrid modes, which induces a power-dependent frequency shift also in the pumped mode. As the pump power increases, this shifts the pumped mode out of resonance, necessitating an adjustment of the pump frequency to maintain resonance condition.

For every couple of hybrid modes, the dependence of the cross-Kerr on the bare SQUID array resonator frequency is obtained by first sweeping the power of the pump tone and then the flux bias applied via the external coil.
For instance, the sequence in \subautoref{fig:cross_kerr_appendix}{b-h} expands the two-tone spectroscopy map for the couple of hybrid modes $(c_{12},c_{p})$ shown in \subautoref{fig:CK}{b}.
For this power sweep, the SQUID array resonator frequency is marked by the red dotted line in \subautoref{fig:cross_kerr_appendix}{a}.

For the sake of simplicity, in the cross-Kerr study, we adopt the notation $c_{p} \in \{c_8, c_9, c_{10}\}$, since the bare modes $b_9 \text{ and } b_{10}$ are included in the numbering, but do not contribute effectively in the model as their coupling is set to zero, $g_9=g_{10}=0$.

Finally, the external and the total dissipation rates of the hybrid modes vary significantly as a function of the frequency of the bare SQUID array resonator.
Therefore, we calibrate the power of the pump tone so that the excitation number in the pumped mode remains $\lesssim 10 $ for any $\wsq$.
In practice, the power of the pump tone is obtained by inverting \autoref{Eq:Photon_number_conversion}, where the frequency and dissipation rates of mode $c_{p}$ as a function of $\wsq$ are obtained from a fit of the $S_{21}$ traces shown in \subautoref{fig:CK}{a} to \autoref{Eq:Notch_coefficient_single_mode_FANO}. 

\subsubsection*{Analysis protocol}
\begin{figure*}
    \includegraphics[width = 1\linewidth ]{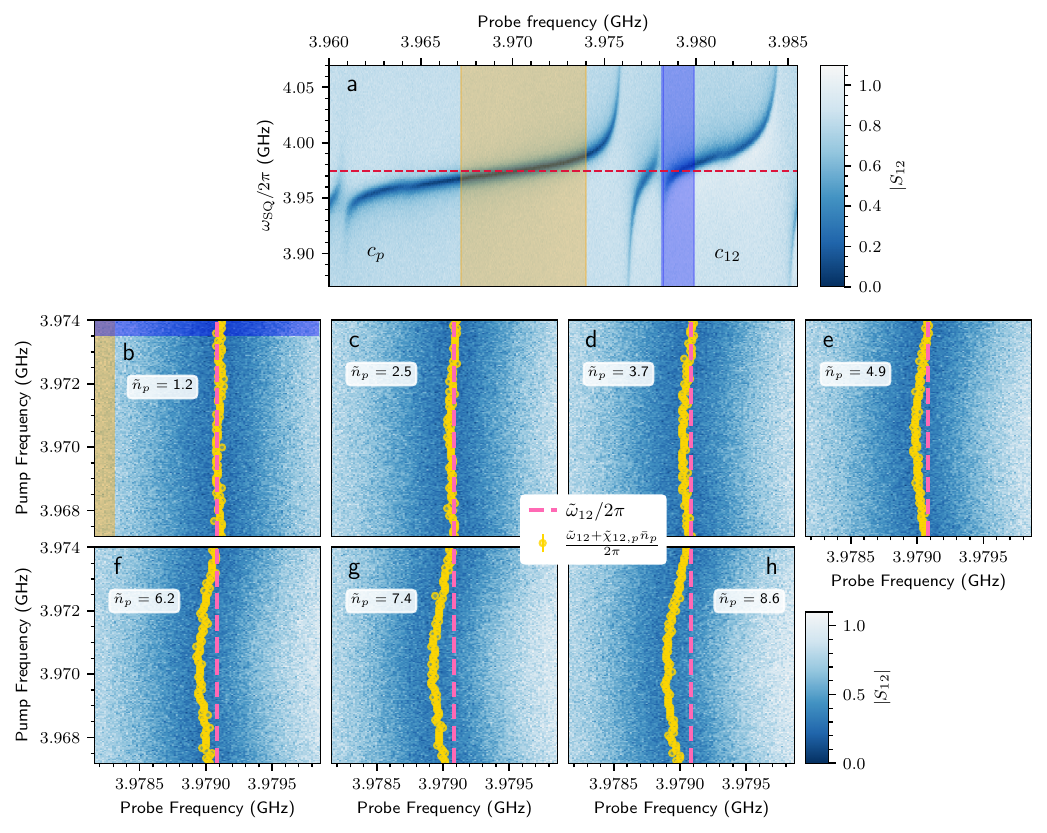}
    \caption{\label{fig:cross_kerr_appendix}
    \textbf{
    Cross-Kerr measurement at one flux value.
    }
    (a) Magnitude of the normalised transmission signal $|S_{21}|$ through the SQUID array resonators feedline as a function of the flux bias threading the SQUIDs loop.
    The dashed red line indicates the bare SQUID array resonator frequency $\wsq$ in (b-h), while shaded regions highlight the frequency range of the pump tone on hybrid modes $c_{p}= \{c_{8}, c_{9}, c_{10}\} $ (orange) and of the probe tone on $c_{12}$ (blue).
    (b-h) Power series of the two-tone spectroscopy maps of hybrid mode $c_{12}$ while pumping the hybrid modes $c_{p}$. 
    The pump power increases from left to right and from top to bottom. 
    The shaded rectangles in (b) correspond to the shaded regions in the top panel. 
    The dashed pink lines indicate the frequency of mode $c_{12}$ in absence of pump tone, while the yellow circles mark the extracted resonance frequency of probed mode $c_{12}$ for each pump frequency.
    }
\end{figure*}

Here, to illustrate the protocol for analysing two-tone spectroscopy maps, we focus on the pair of hybrid modes $(c_{12},c_p)$ .
For each map in \subautoref{fig:cross_kerr_appendix}{b-h}, we extract the frequency of the probed mode $c_{12}$ (yellow circles in \autoref{fig:cross_kerr_appendix} lower panel) as a function of the pump tone frequency, $\tilde{\omega}_{12}(f_{pump})$.
To this end, we fit each $S_{21}$ trace to the notch-type transmission signal \autoref{Eq:Notch_coefficient_single_mode_FANO}. 
However, because of impedance mismatch (or Fano interference), the line shape of some modes is distorted.
This distortion induces a systematic offset between resonance frequency $\omega_r$ obtained from the fit to \autoref{Eq:Notch_coefficient_single_mode_FANO} and the actual minimum in the magnitude of the measured $S_{21}$ trace.
This offset depends on the specific probed mode but is independent of the power applied to the pumped mode $c_p$.
Therefore, to apply a uniform analysis protocol across the entire dataset, we use the minimum of the best-fit curve as the resonance frequency of the probed mode.

Subsequently, we extract the maximum negative displacement, $\Delta \tilde{\omega}^0_{12,p} \coloneqq \Delta \tilde{\omega}^0_{12,p}(\bar{n}_p,\wsq) $
\begin{equation}
    \Delta \tilde{\omega}^0_{12,p} = \underset{f_{pump}}{\min} \left(\tilde{\omega}_{12}(f_{pump})\right) - \tilde{\omega}_{12}^0,
\label{eq:displacement_uncorrected}
\end{equation}
where $\tilde{\omega}^0_{12}/2\pi$ is the resonance frequency of the hybrid mode $c_{12}$ from diagonalisation of the linear model \autoref{Eq:Full_Linear_Hamiltonian_RWA}.
In addition, the number of excitations in the pumped mode ($\bar{n}_p$) is re-calibrated by using the frequency of the pump tone at maximum displacement $f_p^* = \underset{f_{pump}}{\arg\min}\left(\tilde{\omega}_{12}(f_{pump})\right)$.

The displacements, $\Delta \tilde{\omega}^0_{12,p}$, against the excitation number, $\bar{n}_p$, are then fitted to a linear model. 
The slope gives the cross-Kerr coefficient at the specific frequency of the SQUID array resonator considered, $\tilde{\chi}_{i,j}(\wsq)$.
The zero-order approximation for $\tilde{\omega}_{12}^0$ is provided by diagonalising the linear model \autoref{Eq:RWA_hamiltonian_matrix}.
However, the required precision on the pump-off position of the probed modes would be below $10^{-5}$. 
Therefore, in the linear fit we include the intercept as a fitting parameter.
The dashed pink line in \autoref{fig:cross_kerr_appendix} is the pump-off frequency of the probed mode $c_{12}$ obtained from the linear model corrected with the intercept from the fit: $\tilde{\omega}_{12}^{}=\tilde{\omega}_{12}^{0} + \rm offset$. 
By substituting the offset corrected frequency in \autoref{eq:displacement_uncorrected}, we obtain $\Delta \tilde{\omega}_{12,p}$, shown in \subautoref{fig:CK}{c}.

Finally, the uncertainty in the calibration of the line is the dominant source of error in the self- and cross-Kerr measurements, and it acts as a global rescaling of the Kerr coefficients.
Therefore, we estimate its impact on the value of self-Kerr of the bare SQUID array resonator extracted from the fit as, 
\begin{equation}
    \Delta\chi_{SQ, \pm} = \chi_{SQ} \times (10^{\pm \Delta\rm A/10} - 1) \, ,
\label{eq:error_SQUID_SK}
\end{equation}
where $\Delta \rm A$ is the uncertainty in the calibration of the SQUID array resonator feedline. 

\newpage 


%

\end{document}